\pdfoutput=1

\documentclass[useAMS,usenatbib]{mn2e}
\usepackage{graphicx}
\usepackage{amsmath} 
\usepackage{times}
\usepackage{verbatim}
\usepackage[T1]{fontenc}
\usepackage{tabularx}
\usepackage{array}
\usepackage{color}
\usepackage{subfig}

\usepackage{hyperref}
\usepackage{abbreviations}

\newcommand{\Brg}{Br$\gamma$}
\newcommand{\Brd}{Br$\delta$}
\newcommand{\ts}{35 mas pixel$^{-1}$}
\newcommand{\hs}{100 mas pixel$^{-1}$}

\newcolumntype{R}{>{\centering\arraybackslash}X}

\begin{document}

\title[Star clusters in MCG+08-11-002]{Reconstructing merger timelines using star cluster age distributions: the case of MCG+08-11-002}

\author[R. L. Davies, A. M. Medling, et al.]{Rebecca L. Davies$^{1}$\thanks{E-mail:
Rebecca.Davies@anu.edu.au}, Anne M. Medling$^{1}$, Vivian U$^{2}$, Claire E. Max$^{3}$, David Sanders$^{4}$, \newauthor Lisa. J. Kewley$^{1}$ \\ \\
$^{1}$Research School of Astronomy and Astrophysics, Australian National University, Canberra, ACT 2611, Australia \\
$^{2}$Department of Physics and Astronomy, University of California, Riverside, 900 University Avenue, Riverside, CA 92521, USA \\
$^{3}$Department of Astronomy \& Astrophysics, University of California, Santa Cruz, CA 95064, USA \\
$^{4}$ Institute for Astronomy, University of Hawaii, 2680 Woodlawn Dr., Honolulu, HI 96822, USA}

\maketitle

\begin{abstract}
We present near infrared imaging and integral field spectroscopy of the centre of the dusty luminous infrared galaxy merger \mbox{MCG+08-11-002}, taken using the Near InfraRed Camera 2 (NIRC2) and the OH-Suppressing InfraRed Imaging Spectrograph (OSIRIS) on Keck II. We achieve a spatial resolution of \mbox{$\sim$25 pc} in the K band, allowing us to resolve 41 star clusters in the NIRC2 images. We calculate the ages of 22/25 star clusters within the OSIRIS field using the equivalent widths of the \mbox{CO 2.3$\mu$m} absorption feature and the \Brg\ nebular emission line. The star cluster age distribution has a clear peak at ages \mbox{$\la$ 20 Myr}, indicative of current starburst activity associated with the final coalescence of the progenitor galaxies. There is a possible second peak at \mbox{$\sim$65 Myr} which may be a product of the previous close passage of the galaxy nuclei. We fit single and double starburst models to the star cluster age distribution and use Monte Carlo sampling combined with two-sided K-S tests to calculate the probability that the observed data are drawn from each of the best fit distributions. There is a $>$90 per cent chance that the data are drawn from either a single or double starburst star formation history, but stochastic sampling prevents us from distinguishing between the two scenarios. Our analysis of MCG+08-11-002 indicates that star cluster age distributions provide valuable insights into the timelines of galaxy interactions and may therefore play an important role in the future development of precise merger stage classification systems.
\end{abstract}

\section{Introduction}
Luminous (\mbox{$\rm L_{IR} \, \textgreater 10^{11} \, L_{\odot}$}) and Ultra-Luminous (\mbox{$\rm L_{IR} \, \textgreater 10^{12} \, L_{\odot}$}) Infrared Galaxies ((U)LIRGs) appear to be a common but short-lived phase of galaxy evolution, triggered by major mergers of gas-rich spiral galaxies \citep{Armus87, Melnick90, Mihos94, Sanders96}. The fraction of (U)LIRGs undergoing interaction increases with infrared (IR) luminosity, and the local merger fraction surpasses 90\% at the largest IR luminosities \citep{Sanders88, Melnick90, Clements96, Veilleux02, Ishida04, Haan11}. The incidence of (U)LIRGs in galaxy pairs also increases as pair separation decreases \citep{Ellison13}. Galaxy interactions drive strong tidal torques which funnel large amounts of gas towards the centres of merging systems, producing dense gas reservoirs which trigger black hole growth and bursts of star formation. LIRGs are therefore ideal laboratories for studying the build up of stellar mass and the relationship between star formation and AGN activity in galaxies.

Accurate and high temporal resolution merger stage classification systems are required to understand how star formation and AGN activity evolve over the course of galaxy mergers. Morphological properties such as the projected separation of the progenitor nuclei, the presence or absence of tidal features and the length of tidal tails (if present) are commonly used as proxies for merger stage \citep[see e.g.][]{Veilleux02, Yuan10}. However, the evolution of nuclear separation as a function of merger stage 1) is non monotonic, and 2) is dependent on the initial orbital parameters of the progenitor systems \citep[e.g.][]{Barnes92, Privon13}. The ability to detect tidal features at a given merger stage is very dependent on the mass ratio, gas properties, bulge-to-disk ratios, orbits and dust contents of the progenitor galaxies, as well as the depth and wavelength of observations \citep[e.g.][]{Schawinski10, Kartaltepe12, Privon13, Snyder15}. Quantitative measures of morphology such as the Gini, $M_{20}$ and Asymmetry metrics are able to distinguish mergers at first passage or final coalescence from normal galaxies, but do not reliably identify galaxies at other stages of merging \citep{Lotz08}. Further information is required to determine the merger stages of galaxies more precisely.

Matching the observed properties of merging systems to the properties of galaxies in merger simulations offers direct insights into merger timelines but is very computationally expensive due to the size of the parameter space that must be explored. Fortunately, the release of libraries of galaxy merger simulations has significantly improved their accessibility within the wider astronomical community. The GalMer library provides mock images, spectra and datacubes of merging systems at \mbox{50 Myr} intervals for a range of initial conditions \citep{Chilingarian10}. Improvements in the efficiency of N-body dynamical models have also made it possible to construct ensembles of merger models using packages such as \textsc{Identikit} \citep{Barnes09, Privon13}. However, significant issues remain in attempting to break the degeneracies within multi-dimensional parameter spaces, particularly when the observed morphologies of galaxies are so dependent on the depth and wavelength of observations, viewing angle and dust extinction. 

Star formation rates vary strongly during the course of galaxy mergers and may offer valuable insights into merger stage. Simulations of gas-rich merging systems show strong peaks in star formation rate corresponding to close passages between the galaxy nuclei \citep[e.g.][]{Mihos96, DiMatteo05, DiMatteo07, Hopkins13, Renaud14}. Observations have confirmed that galaxies with companions at projected separations of \mbox{$\la$ 30 kpc} have significantly larger star formation rates than isolated galaxies at the same stellar mass and redshift \citep{Barton00, Lambas03, Ellison08, FreedmanWoods10, Scudder12, Patton13, Ellison13}. The star formation histories of (U)LIRGs are therefore important tracers of their interaction histories.

High resolution \textit{Hubble Space Telescope} (HST) Advanced Camera for Surveys (ACS) imaging has facilitated demographic studies of the nuclear star cluster populations of many merging systems. The presence of distinct young (\mbox{$\la$ 20 Myr}) and intermediate age (\mbox{$\sim$ 100 - 500 Myr}) star cluster populations in late stage mergers such as the Antennae galaxies (\mbox{NGC 4038/4039}; \citealt{Whitmore99}), \mbox{Arp 220} (\citealt{Wilson06}), the Mice galaxies (\mbox{NGC 4676 A/B}; \citealt{Chien07}) and \mbox{NGC 7252} \citep{Miller97} is suggestive of individual starburst events triggered during close pericentre passages of the progenitor galaxies. However, optical studies of other late stage mergers such as \mbox{NGC 6240} \citep{Pasquali03} and \mbox{NGC 7673} \citep{Homeier02} reveal only young star clusters, indicating that older star clusters from the first pericentre passage are either rare or undetected. In contrast, the lack of very young (\mbox{$\leq$10 Myr}) star clusters in the tidal tails of \mbox{NGC 520}, \mbox{NGC 2623} and \mbox{NGC 3256} suggests that the cold gas has already been consumed by previous star formation \citep{Mulia15}.

Unfortunately, resolved HST/ACS imaging can only detect star clusters in regions with relatively little dust obscuration. Prolific star formation and AGN activity surrounding the final coalescence phase of merging galaxies produces a large amount of obscuring dust, (with the optical depth at 550nm reaching 30 towards the nuclear regions of some ULIRGs), which absorbs the optical emission of the young star clusters and re-emits it in the infrared \citep[e.g.][]{Veilleux95, Sanders96, Desai07}. Investigating the rate and history of star formation in heavily obscured regions is pivotal for constructing representative star cluster age distributions of merging systems, and may provide new insights into the building of stellar mass during the most rapid periods of galaxy evolution. The recent development of advanced adaptive optics (AO) systems has made it possible for infrared observations from ground-based 8-10m class telescopes to achieve similar (or better) spatial resolution to optical images taken with HST/ACS. This capability has led to the identification of many previously unseen star clusters in dusty (U)LIRGs, including the discovery of intermediate age star clusters in \mbox{NGC 6240} at a redshift of \mbox{z = 0.024} \citep[e.g.][]{Max05, Pollack07}.

MCG+08-11-002 (hereafter MCG08; \mbox{z = 0.0198}, \mbox{$d$ = 86 Mpc}, \mbox{1 arcsec = 417 pc}) is a LIRG (\mbox{$\rm L_{IR} = 10^{11.46} L_{\odot}$}; \citealt{Armus09}) and a late stage merger, with two distinct nuclei separated by $\sim$0.32$\arcsec$ (\mbox{133 pc}). The galaxy is drawn from the Great Observatories All-sky LIRG Survey (GOALS), a flux-limited survey of the brightest infrared galaxies at z $\textless$ 0.1. Prominent \Brg\ and silicate emission signatures reveal the presence of young star clusters in the nuclear region \citep{Diaz-Santos10, Diaz-Santos11, Medling14}. However, MCG08 is undetected in GALEX far-ultraviolet (FUV) observations and its near-UV (NUV) flux density (\mbox{6.8$\times$10$^{-16}$ erg s$^{-1}$ cm$^{-2}$ \AA$^{-1}$}) is more than 4 standard deviations below the average NUV flux density within the GOALS sample (\mbox{1.3$\pm$0.3 $\times$ 10$^{-14}$ erg s$^{-1}$ cm$^{-2}$ \AA$^{-1}$}). The weakness of the UV flux suggests that the ionizing radiation from the young nuclear star clusters within MCG08 is heavily dust attenuated \citep{Howell10}. The mass of the central black hole is in the range \mbox{1.2 $\times 10^{7}$} - \mbox{1.3 $\rm \times 10^{9}$} $M_{\odot}$ \citep{Medling15} but no active galactic nucleus (AGN) has been detected \citep{Petric11}.

MCG08 has been imaged by HST/ACS in the F814W and F435W filters as part of the GOALS survey (Cycle 14, Evans et al., Program \#10592). The images have a field of view (FOV) of \mbox{216$\arcsec$ $\times$ 216$\arcsec$} at a spatial sampling of \mbox{0.05 arcsec pixel$^{-1}$}. However, very few of nuclear star clusters are detected at visible wavelengths due to the strong dust attenuation. In order to uncover the buried star clusters we have obtained high resolution near infrared (NIR) imaging and integral field spectroscopy using the Near Infrared Camera 2 (NIRC2) and the OH-Suppressing Infra-Red Imaging Spectrograph (OSIRIS) on Keck II (aided by the Keck laser guide star AO system). We describe our observations, data processing and star cluster identification in Section \ref{sec:observations}. We use spectroscopic indices to constrain the ages of the nuclear star clusters in Section \ref{sec:spectro_constraints}. We discuss the completeness of our sample and compare the observed star cluster age distribution with models for different star formation histories in Section \ref{sec:stats}. We summarise our results and present our conclusions in Section \ref{sec:conclusions}.

\begin{figure}
\centerline{\includegraphics[scale=0.85]{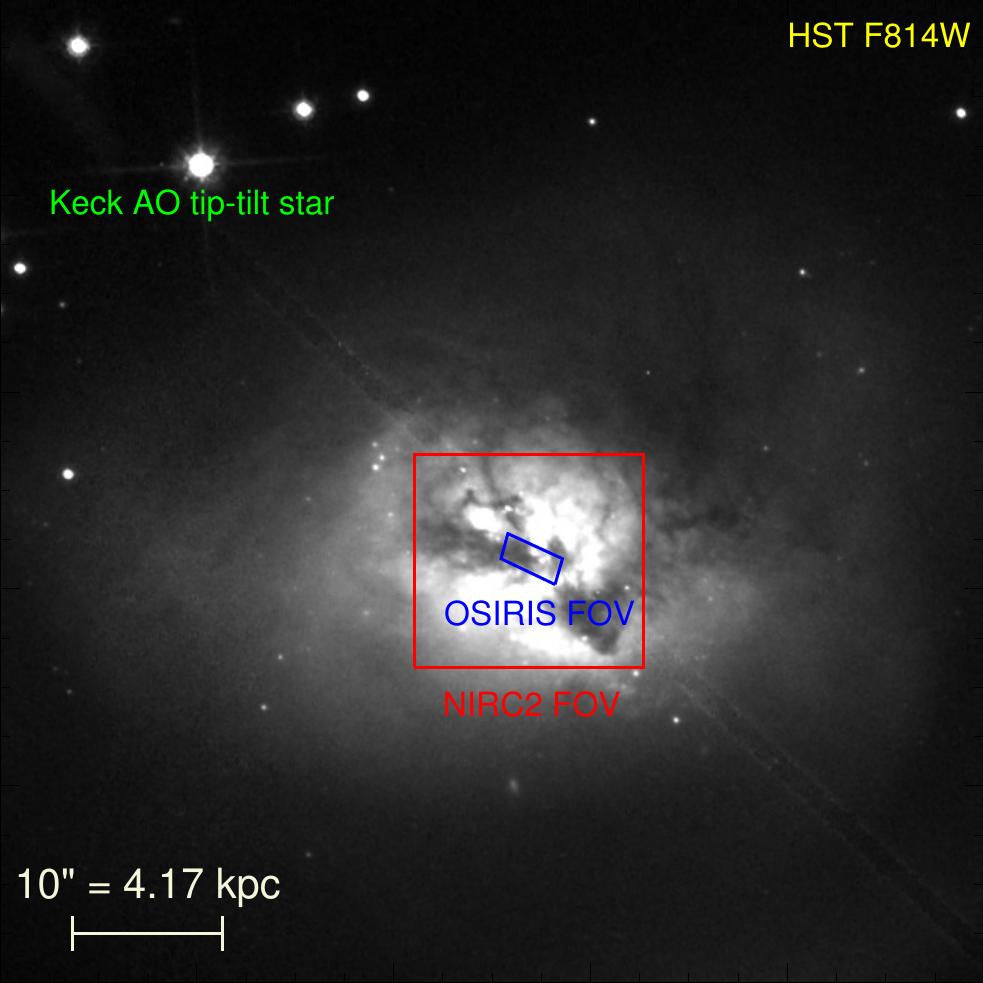}}
\caption{Footprints of the Keck NIRC2 (red) and OSIRIS \hs\ (blue) observations overlaid on the archival HST/ACS F814W image of MCG08. The sizes of the boxes match the FOV listings from Table \ref{table:exposures}. The tip-tilt star used with the Keck adaptive optics system is located in the top left hand corner of the Figure.}\label{Fig:footprint}
\end{figure}

Throughout this paper, we adopt cosmological parameters \mbox{$\rm H_0 = 70.5 \, km \, s^{-1} \, Mpc^{-1}$}, \mbox{$\Omega_M$ = 0.27} and \mbox{$\Omega_{\Lambda}$ = 0.73} based on the 5-year \textit{Wilkinson Microwave Anisotropy Probe} (WMAP) results \citep{Hinshaw09}. 

\section{Observations and data processing}
\label{sec:observations}

\begin{table*}
    \caption{Details of Keck NIRC2 and OSIRIS observations of MCG+08-11-002. The field of view listings for the OSIRIS observations are the final high signal-to-noise regions used for emission line fitting after individual science observations were mosaicked.}
    \begin{tabularx}{\textwidth}{RRRRRRRRR}
    \hline 
    \hline
   	Instrument & Filter & Wavelength coverage ($\mu$m) & \multicolumn{2}{c}{FOV of final frames} & Arcsec pixel$^{-1}$ & UT Date(s) (YYMMDD) & Science exp time (s) & Sky exp time (s) \\ \hline
	OSIRIS & Kcb (night 1) & 1.965 - 2.381 & \multicolumn{2}{c}{1.5$\arcsec$ $\times$ 3.2$\arcsec$} & 0.1 & 110110 & 2400 & 1200 \\ 
	'' & Kcb (night 2) & '' & \multicolumn{2}{c}{''} & '' & 120102 & 1200 & 600 \\ 
	'' & Kbb & '' & \multicolumn{2}{c}{0.74$\arcsec$ $\times$ 1.79$\arcsec$} & 0.035 & 120102 & 3600 & 1800 \\ \hline
	NIRC2 & Kp & 1.948 - 2.299 & \multicolumn{2}{c}{10$\arcsec$ $\times$ 10$\arcsec$} & 0.01 & 121223 & 660 & - \\
	'' & J (night 1) & 1.166 - 1.330 & \multicolumn{2}{c}{''} & '' & 121223 & 360 & - \\
	'' & J (night 2) & '' & \multicolumn{2}{c}{''} & '' & 130206 & 780 & - \\ \hline
	\end{tabularx}
    \label{table:exposures}
\end{table*}

\subsection{Keck Observations}
We observed the nuclear region of MCG08 using both NIRC2 and OSIRIS \citep{Larkin06} on the W. M. Keck II telescope between January 2011 and February 2013. The wavelength coverage, field of view, exposure dates and exposure times of our Keck observations are summarised in Table \ref{table:exposures}. The footprints of the NIRC2 and OSIRIS observations of MCG08 are shown in Figure \ref{Fig:footprint}, overlaid on the archival HST/ACS F814W image. 

Our NIRC2 observations were taken using the Kp and J broadband filters, with a FOV of \mbox{10$\arcsec \times$ 10$\arcsec$} at a spatial sampling of \mbox{0.01 arcsec pixel$^{-1}$} (corresponding to \mbox{4.17 pc} at the redshift of MCG08). We observed two standard stars from the UKIRT Faint Standards List \citep{Hawarden01} for flux calibration purposes. We reduced our NIRC2 data using the reduction pipeline of \citet{Do13}, which subtracts sky emission and dark current, flat-fields the images, removes cosmic rays and bad pixels, and corrects for atmospheric and instrumental distortion using the models determined by \citet{Lu08} and \citet{Yelda10}. Individual reduced frames were shifted and median-combined by hand to produce the final image. 

Our OSIRIS observations were taken with both the \hs\ and \ts\ plate scales, using the Kcb and Kbb filters with spatial coverages of {1.5$\arcsec$ $\times$ 3.2$\arcsec$} and {0.74$\arcsec$ $\times$ 1.79$\arcsec$} respectively. Observing MCG08 at both plate scales provides us with the spatial coverage to probe the outer regions of the \hs\ field as well as the spatial resolution to probe the innermost region of the galaxy in great detail. Our observations were carried out in sets of three ten minute exposures, employing an object-sky-object dither pattern. The OSIRIS data were reduced using the OSIRIS Data Reduction Pipeline\footnote{Available at \href{http://www2.keck.hawaii.edu/inst/osiris/tools/}{http://www2.keck.hawaii.edu/inst/osiris/tools/}} version 2.3 which subtracts sky emission, adjusts channel levels, removes crosstalk, identifies glitches in the data, removes cosmic rays and then extracts a 1D spectrum for each spaxel in the integral field data cube, corrects for atmospheric dispersion and telluric absorption and mosaics frames together. We isolated the central high signal-to-noise region of the mosaicked data cubes and then fit the stellar continuum and \Brg\ and \Brd\ emission lines using the method described in \citet{U13}. All references to the FOV of the OSIRIS observations refer to the region over which this fitting was performed (see Table \ref{table:exposures}). The emission lines were assumed to have single-component Gaussian profiles with the same velocity and velocity dispersion. These OSIRIS observations were first presented in \citet{Medling14}, in which a detailed explanation of the data reduction and emission line fitting procedures can be found.

Both NIRC2 and OSIRIS sit behind the Keck Observatory Laser Guide Star (LGS) AO system \citep{Wizinowich00, vanDam04, Wizinowich06, vanDam06} which provides near diffraction limited observations in the NIR. The Keck AO system uses a pulsed laser tuned to the Sodium $D_2$ transition which excites atoms in the sodium layer (at an altitude of \mbox{$\sim$95 km}) and causes spontaneous emission. A Shack-Hartmann wavefront sensor (WFS) monitors the light from the LGS to measure wavefront distortions caused by atmopsheric turbulence. The error signal from the WFS is sent to a deformable mirror which corrects the wavefront distortions in real time. A tip tilt star is also monitored to correct for image motion. The tip tilt star used for our observations has an R-band magnitude of 16.4 and lies 17.6 arcsec from the centre of MCG08 - well within 75 arcsec isokinetic angle (angular distance from the tip-tilt star at which the Strehl ratio will be reduced by 1/$e$) at Mauna Kea \citep{vanDam06}. The typical Strehl ratio achieved by the AO system using a tip tilt star with \mbox{$R \sim$ 17} is 0.25 in Kp band and 0.18 in J band\footnote{\href{https://www2.keck.hawaii.edu/optics/lgsao/performance.html}{https://www2.keck.hawaii.edu/optics/lgsao/performance.html}}. The PSF of the science images, estimated from simultaneous observations of the tip-tilt star, has a FWHM of \mbox{$\sim$ 60 mas} in the NIRC2 Kp band, \mbox{$\sim$ 70 mas} in the NIRC2 J band and \mbox{$\sim$ 90 mas} in the OSIRIS Kbb/Kcb filters. (A more detailed discussion of the NIRC2 PSF characteristics can be found in the \hyperref[subsec:psf_phot]{Appendix}). The PSF is narrower in the Kp band than the J band due to the increased quality of the AO correction at longer wavelengths.

\begin{figure*}
\captionsetup[subfigure]{labelformat=empty}
\subfloat[]{\includegraphics[scale=1.2, clip = true, trim = 0 305 290 0]{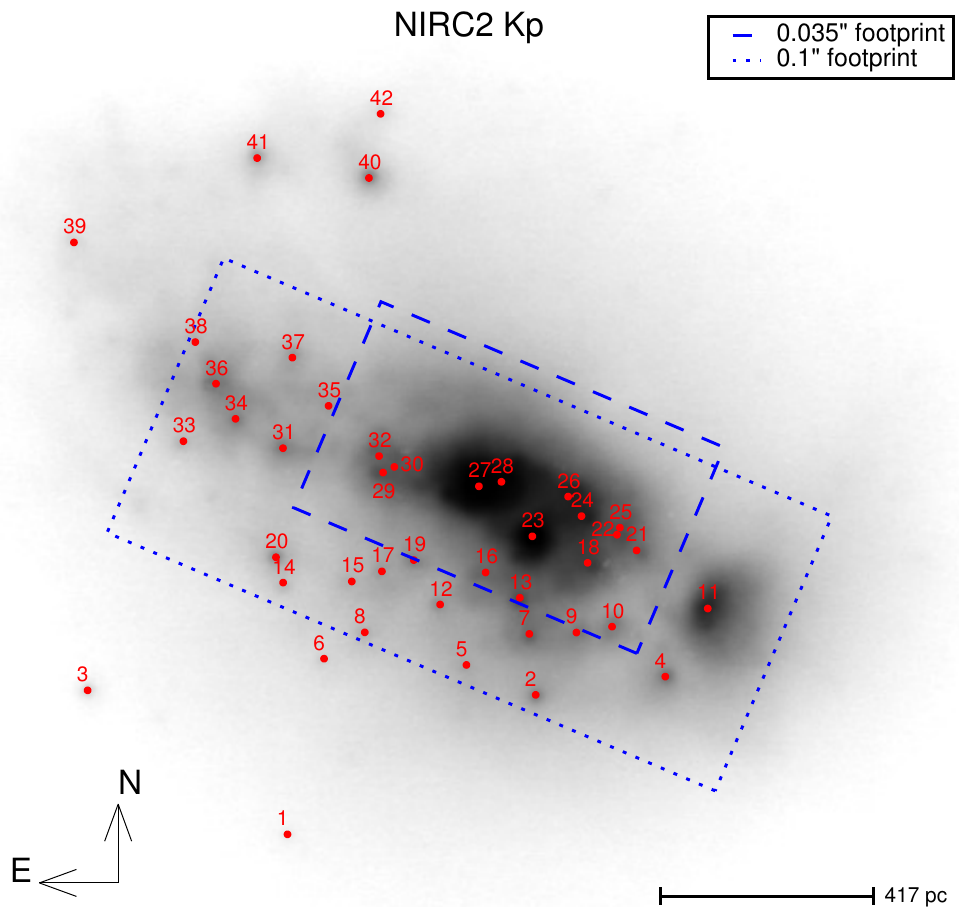}} \\
\subfloat[]{\includegraphics[scale=0.65,clip = true, trim = 0 0 50 0]{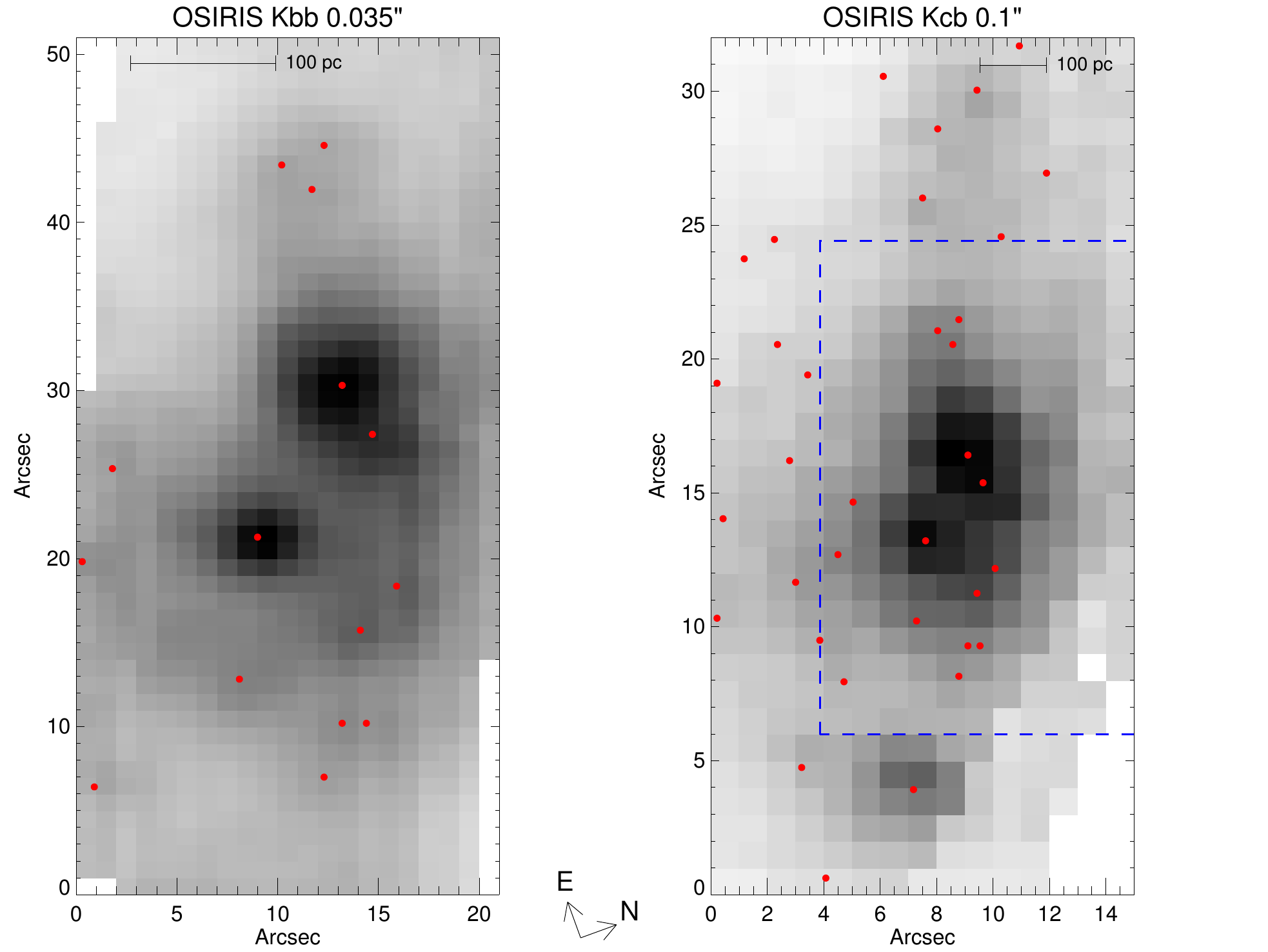}} 
\caption{(Top) NIRC2 Kp band image of the central \mbox{$\sim$ 4$\arcsec$ $\times$ 4$\arcsec$} of MCG08. Red dots indicate the locations of detected star clusters. The numbers assigned to each of the clusters act as their identifiers for the remainder of this paper. The dotted and dashed rectangles show the high signal-to-noise regions of the OSIRIS \hs\ and \ts\ observations respectively. (Bottom) pseudo-continuum images constructed from the OSIRIS \ts\ and \hs\ data respectively. Red dots show the cluster locations calculated by rotating and aligning the OSIRIS pseudo-continuum images to the NIRC2 image.}\label{Fig:alignment}
\end{figure*} 

\begin{figure*}
\centerline{\includegraphics[scale=1.7,clip=true,trim=0 200 0 0]{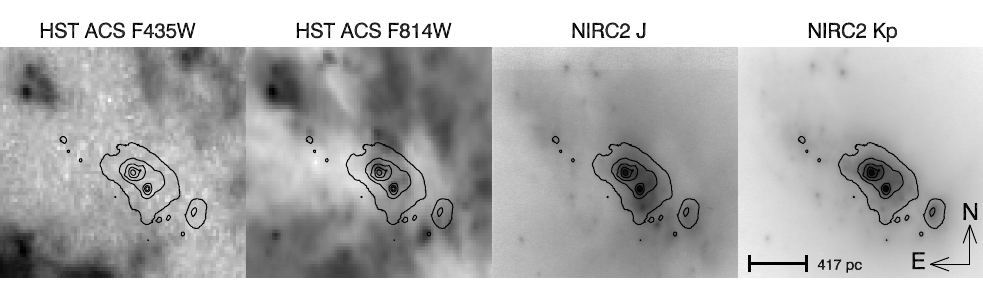}}
\caption{The central region of MCG08 in the F435W, F814W, J and Kp bands. The infrared emission of the galaxy is strongly concentrated in the nuclear region where many star clusters are visible. However, large amounts of dust prevent the majority of these clusters from being detectable in the F814W image (with the exception of the most prominent central clusters) and prevent all clusters from being detectable in the F435W image.}\label{Fig:multiband_images}
\end{figure*}

\subsection{Star cluster identification and image alignment}
\label{subsec:identification}
We identify star clusters by applying the FIND procedure in IDL (an adapted version of the IRAF procedure DAOFIND) to the NIRC2 Kp band image. The code identifies brightness perturbations based on the intensity of each pixel relative to the background, the expected FWHM of the sources, and their 2D sharpness and roundness. We set the cluster detection limit to be 2$\sigma$ above the background, the estimated FWHM of the clusters to be 10 pixels (0.1 arcsec or 41.7 pc), and the approximate light distribution from the clusters to be geometrically round with a Gaussian radial intensity profile. Varying these parameters changes only the number of detected star clusters and not the assigned cluster centroids. All of the detected clusters correspond to visually identifiable brightness perturbations, confirming that the 2$\sigma$ detection limit is sufficient to avoid spurious detections. The top panel of Figure \ref{Fig:alignment} shows the NIRC2 Kp band images of MCG08, with red dots indicating the centroids of the 41 star clusters identified by FIND. The centroids match well with visually identified flux peaks. 

The OSIRIS and NIRC2 observations are aligned by comparing the NIRC2 Kp band image (\mbox{1.948 - 2.299$\mu$m}) with pseudo-continuum images for each of the OSIRIS cubes, created by summing the flux over all wavelength channels (\mbox{1.965 - 2.381$\mu$m}) in each pixel individually. The NIRC2 image is rotated by 70$^\circ$ to match the position angle of the OSIRIS observations and we correct linear offsets in the $x$ and $y$ directions by eye. The dotted and dashed rectangles in the top panel of Figure \ref{Fig:alignment} show the FOV of the OSIRIS \hs\ and \ts\ observations respectively relative to the NIRC2 image. The bottom panels of Figure \ref{Fig:alignment} show the pseudo-continuum images constructed from the \ts\ and \hs\ OSIRIS data respectively. Red dots indicate the locations of the cluster centroids calculated using the alignment transform applied to the NIRC2 image. The cluster centroids are well aligned with the flux peaks in the pseudo-continuum images, indicating that the alignment has been successful.

\subsection{Optical-NIR spectral energy distributions}
\label{subsec:SEDs}
The J band and F814W images are aligned with the NIRC2 Kp band image by applying the FIND algorithm and comparing the centroids of the three brightest clusters. Extreme dust attenuation prevents any of the star clusters being detectable as point-like sources in the F435W image (see Figure \ref{Fig:multiband_images}), so we apply the offsets calculated for the F814W image. The aligned images are used to investigate the optical-NIR spectral energy distributions (SEDs) of the 41 Kp band detected clusters in our sample. We use the 2D light distributions of the tip-tilt star (in the NIRC2 filters) and isolated stars (in the HST images) to determine the PSF in each filter, and extract the flux of each cluster in each filter using PSF photometry. The majority of the derived F814W and F435W fluxes are strict upper limits due to the presence of obscuring dust which prevents the star clusters from being detectable as point-like sources at visible wavelengths. We calculate SED model grids using the Flexible Stellar Population Synthesis (FSPS) code \citep{Conroy10}, and compare the \mbox{J-Kp} color of each cluster with each model in the grids to derive age probability distribution functions (PDFs). Unfortunately our F814W and F435W magnitude limits are not sufficient to break the age-optical depth degeneracy and therefore the ages of the star clusters remain unconstrained. We therefore use spectroscopic information to further investigate the ages of the clusters. A full description of our PSF characterisation, magnitude calculations and model calculations is included in the \hyperref[sec:appendix]{Appendix} for completeness.

\section{Spectroscopic constraints}
\label{sec:spectro_constraints}

\subsection{CO 2.3$\mu$m and \Brg\ equivalent width measurements}
\label{subsec:EWs}
The integrated spectra of stellar populations are shaped primarily by their effective temperatures and therefore their ages. The hottest, most massive (O-B) stars produce strong ultraviolet continuum and absorption features, intermediate (A-G) stars produce prominent spectral features at visible wavelengths, and strong infrared molecular absorption is produced by the coolest (K-M) stars. Hydrogen recombination lines are pronounced features of the integrated spectra of \mbox{$\sim$Myr} old stellar populations.

We probe the ages of the star clusters in our sample by measuring the equivalent widths of two prominent stellar spectral features lying within the wavelength coverage of our OSIRIS data - the CO 2.3$\mu$m absorption feature and the \Brg\ nebular emission line. The CO 2.3$\mu$m absorption feature is strongest in K-M giants and supergiants (dominant in \mbox{$\sim$10 Myr} old stellar populations) but becomes increasingly saturated as the stellar temperature decreases (and therefore as age increases; \citealt{Origlia93}). Nebular \Brg\ emission is excited by emission from massive stars. The equivalent width of the \Brg\ emission line ($W_{Br\gamma}$) exceeds 100 in \mbox{H II} regions younger than 5 Myr but drops rapidly as age increases, becoming virtually undetectable in stellar populations older than \mbox{35 Myr} (see e.g. \citealt{Leitherer99}). The combination of $W_{CO}$ and $W_{Br\gamma}$ allows us to probe star clusters with ages spanning from \mbox{1 Myr} to \mbox{1 Gyr} (see Section \ref{subsec:ages}).

\begin{figure*}
\captionsetup[subfigure]{labelformat=empty}
\subfloat[]{\includegraphics[scale = 1.05, clip = true, trim = 0 140 240 0]{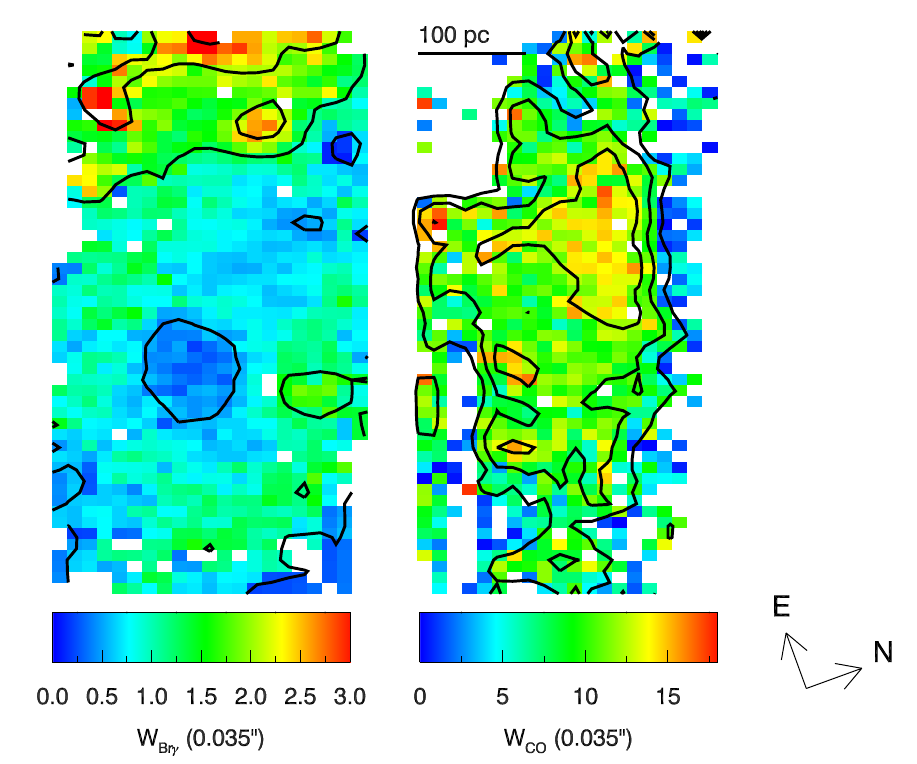}}
\subfloat[]{\includegraphics[scale = 1.05, clip = true, trim = 0 140 295 10]{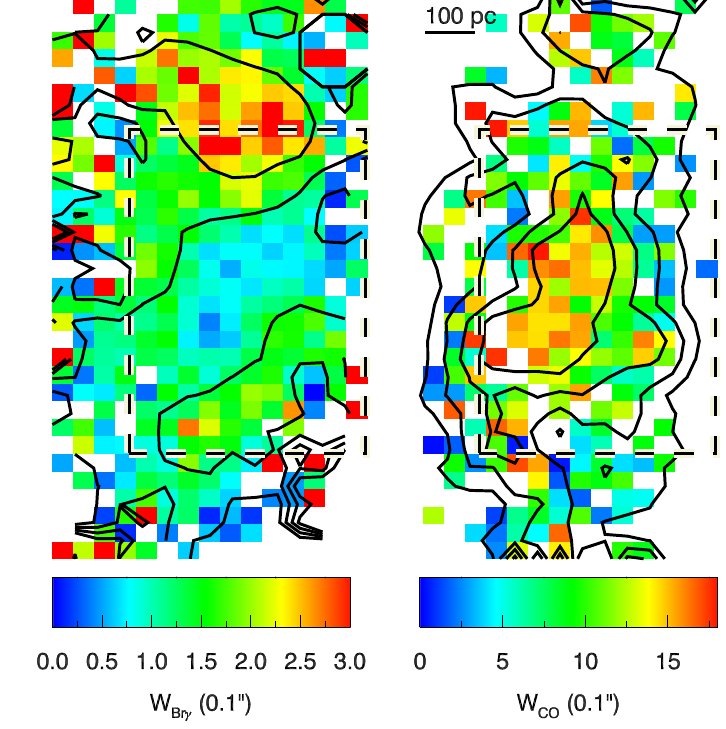}} 
\caption{Maps with contours overlaid of (left to right) the equivalent width of the \Brg\ emission line and the equivalent width of the CO 2.3$\mu$m absorption feature from the \ts\ data, and the same from the \hs\ data. The contour levels are consistent between plate scales for each of the indices. Dashed rectangles on the \hs\ plots indicate the coverage of the \ts\ observations.}\label{Fig:EW_maps}
\end{figure*}

\begin{figure*}
\centerline{\includegraphics[scale=1,clip=true,trim=0 150 0 0]{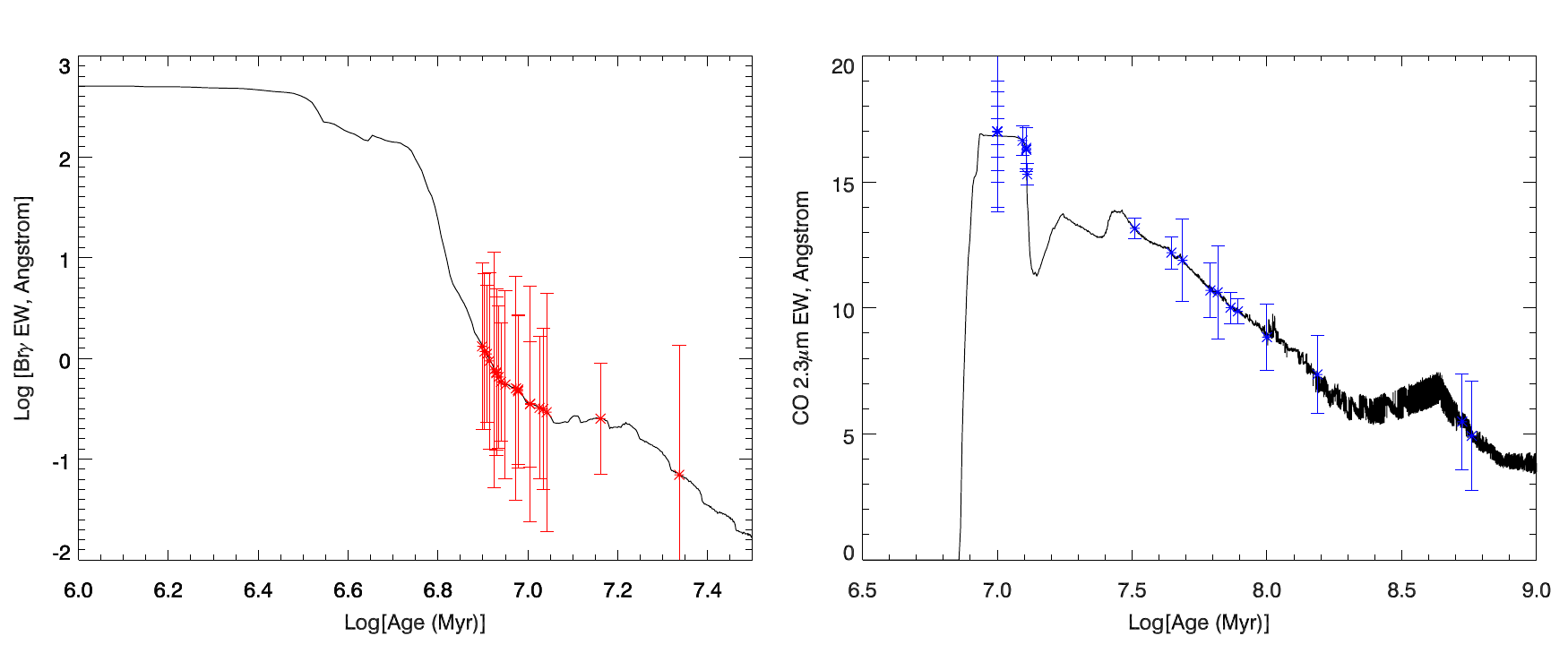}}
\caption{Starburst99 models for the (left) \Brg\ and (right) CO 2.3$\mu$m equivalent widths as a function of star cluster age, assuming a Salpeter IMF and solar metallicity (Z = 0.020). Coloured data points illustrate the correspondence between equivalent width measurements and age estimates for individual star clusters in MCG08.}\label{Fig:sb99_CO}
\end{figure*} 

Figure \ref{Fig:EW_maps} shows maps of (left to right) $W_{Br\gamma}$ and $W_{CO}$ for the 35 mas pixel$^{-1}$ data with contours of the integrated continuum overlaid, and the same for the 100 mas pixel$^{-1}$ data. The equivalent widths are calculated using the definitions adopted by \citet{Leitherer99}. The average $W_{CO}$ values are 8.9\AA\ and 10.0\AA\ for the \ts\ and \hs\ respectively, and the average $W_{Br\gamma}$ values are 1.05\AA\ and 1.4\AA. These values are indicative of a \mbox{$\sim$5 - 100 Myr} old stellar population \citep{Leitherer99}. The average $W_{CO}$ values appear to decrease towards the outer regions of the \ts\ field. In contrast, the $W_{Br\gamma}$ values are smallest near the centre of the galaxy and peak strongly in the upper (eastern) regions of the OSIRIS fields. The clear difference in the morphology of the $W_{CO}$ and $W_{Br\gamma}$ maps highlights their differing sensitivities and complementary nature.  

\begin{figure*}
\centerline{\includegraphics[scale=1]{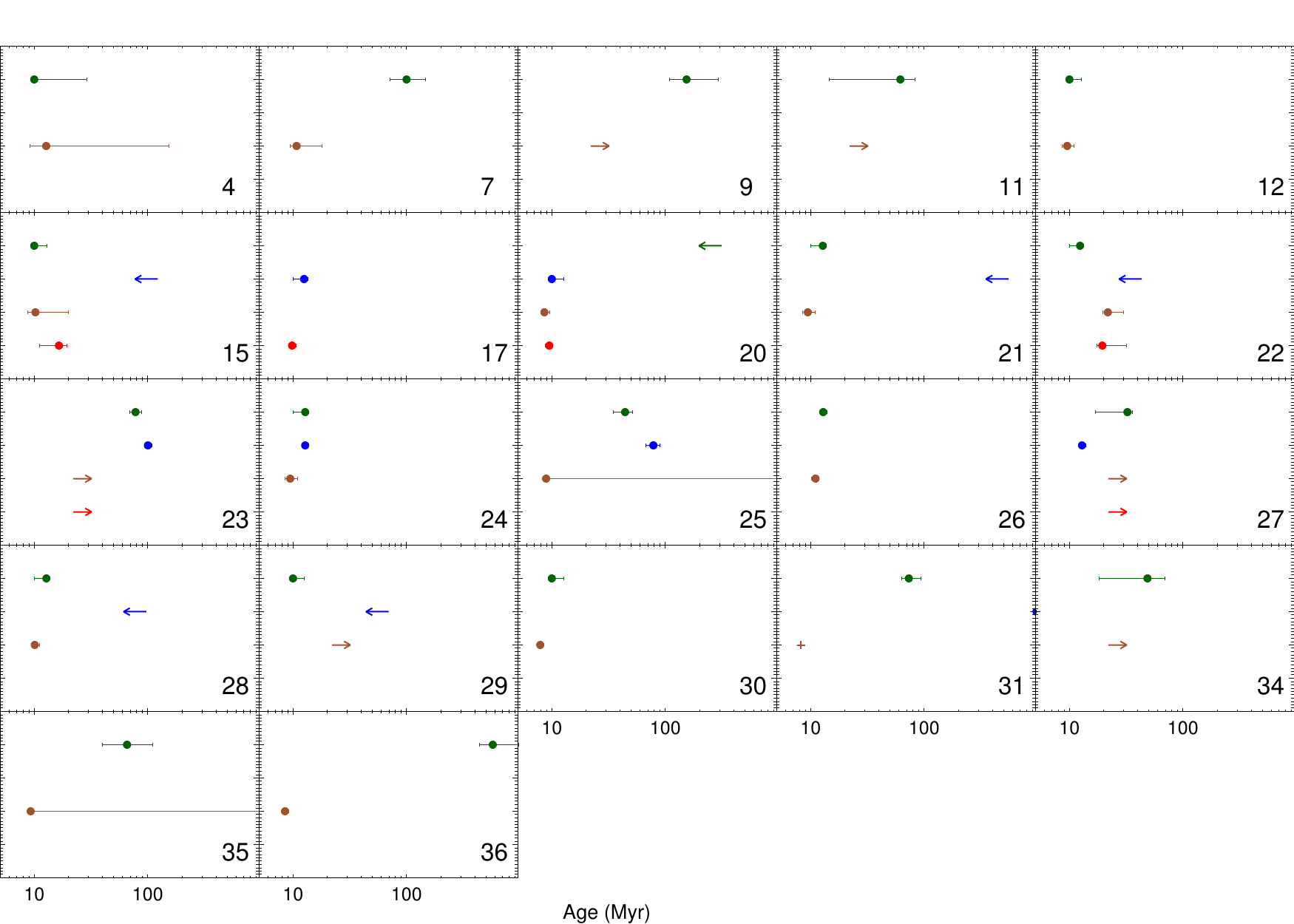}} 
\caption{Age estimates from (green) $W_{CO}$ \hs, (blue) $W_{CO}$ \ts, (brown) $W_{Br\gamma}$ \hs and (red) $W_{Br\gamma}$ \ts\ data for each of the clusters. Numbers in the bottom left corner of each panel correspond to the cluster identifiers in Figure 2 and Tables \ref{Table:table_age} and \ref{Table:appendix}. Arrows indicate age limits deduced from raw equivalent width values (no background subtraction) or detection of only one spectral feature in a particular cluster. The `+' sign in the panel for cluster 31 indicates that the \Brg\ measurement may be unreliable due to contamination from nearby clusters (28 and 29). Individual age estimates are consistent with one another within the errors for 68 per cent (15/22) of clusters and within twice the size of the errors for 82 per cent (18/22) of clusters.}\label{Fig:age_measurements}
\end{figure*}

The spectrum of each spaxel of the OSIRIS data cubes is a combination of emission and absorption from young stellar populations as well as background sources such as old stellar populations, hot dust and/or an AGN. (There is no evidence for an AGN in the nucleus of MCG08. However, if an AGN were present, it would only contribute significantly to the continuum emission very close to the kinematic centre of the system.) Emission from background sources will dilute the CO 2.3$\mu$m absorption feature, such that our measured $W_{CO}$ values are lower limits to the true values and the resulting age estimates are upper limits. We construct an average `background' spectrum by averaging the spectra of all spaxels in the outermost rows and columns of each data cube, weighted by the signal-to-noise of the (K band) continuum. We exclude any spaxels which lie within 0.5 arcsec of a cluster centroid. The background spectrum is then subtracted from all spaxels in the data cube and the $W_{CO}$ and $W_{Br\gamma}$ measurements for each cluster are extracted directly from the spaxel containing the cluster centroid in each plate scale. 61 per cent (25/41) of the clusters lie within the FOV of the \hs\ data, and 13 of these (32 percent of the clusters and 52 percent of the spectroscopic sample) also lie within the FOV of the \ts\ data. We do not impose any signal-to-noise limits on our equivalent width measurements, ensuring that any excess \Brg\ emission or CO absorption above the background level will be detected.

The purpose of the background subtraction is to remove contamination diluting the \mbox{CO 2.3$\mu$m} absorption feature. Therefore, $W_{CO}$ should increase after background subtraction if the derived background spectrum adequately characterises the background emission across the entire OSIRIS FOV. We test the success of the background subtraction algorithm by comparing the raw and background-subtracted $W_{CO}$ values for each cluster. In 19/25 cases, $W_{CO}$ (calculated from the \hs\ data and the \ts\ data when available) increases after background subtraction (as expected), but in 6/25 cases it decreases. For these clusters we retain the raw $W_{CO}$ values and derive upper limits on their ages.

\begin{table*}
	\caption{Centroid locations (in pixels for the NIRC2 Kp band image) and age estimates (from CO \ts\ and \hs\ data and \Brg\ \ts\ and \hs\ data) for each of the 41 star clusters in our sample. Clusters which do not have any listed age estimates are not covered by our OSIRIS observations.}
	\begin{tabularx}{\textwidth}{RRRRRRR}
    \hline
    \hline
    Identifier & \multicolumn{2}{c}{Centroids (pixels)} & \multicolumn{2}{c}{CO Age (Myr)} & \multicolumn{2}{c}{\Brg\ Age (Myr)} \\ \hline
   	~ & $x$ & $y$ & \ts\ & \hs\ & \ts\ & \hs\ \\ \hline 
1 & 484.9 & 419 & -$^{ }_{ }$ & -$^{ }_{ }$ & -$^{ }_{ }$ & -$^{ }_{ }$  \\
2 & 601.4 & 484.4 & -$^{ }_{ }$ & -$^{ }_{ }$ & -$^{ }_{ }$ & -$^{ }_{ }$  \\
3 & 391.1 & 486.6 & -$^{ }_{ }$ & -$^{}_{}$ & -$^{ }_{ }$ & -$^{ }_{ }$  \\
4 & 662.2 & 493 & -$^{ }_{ }$ & 10$^{+19}_{-0}$ & -$^{ }_{ }$ & 16$^{+187}_{-5}$  \\
5 & 568.9 & 498.5 & -$^{ }_{ }$ & -$^{ }_{ }$ & -$^{ }_{ }$ & -$^{ }_{ }$  \\
6 & 502.1 & 501.4 & -$^{ }_{ }$ & -$^{ }_{ }$ & -$^{ }_{ }$ & -$^{ }_{ }$  \\
7 & 598.4 & 513 & -$^{ }_{ }$ & 100$^{+47}_{-29}$ & -$^{ }_{ }$ & 14.2$^{+9}_{-2}$  \\
8 & 521.2 & 513.8 & -$^{ }_{ }$ & -$^{ }_{ }$ & -$^{ }_{ }$ & -$^{ }_{ }$  \\
9 & 620.5 & 513.7 & -$^{ }_{ }$ & 154$^{+138}_{-45}$ & -$^{ }_{ }$ & > 32$^{ }_{ }$  \\
10 & 637.2 & 516.4 & -$^{ }_{ }$ & -$^{ }_{ }$ & -$^{ }_{ }$ & -$^{ }_{ }$  \\
11 & 682.1 & 525 & -$^{ }_{ }$ & 62$^{+21}_{-47}$ & -$^{ }_{ }$ & > 32$^{ }_{ }$  \\
12 & 593.9 & 530 & -$^{ }_{ }$ & 10$^{+3}_{-0}$ & -$^{ }_{ }$ & 12$^{+2}_{-1}$  \\
13 & 482.9 & 537.1 & -$^{ }_{ }$ & -$^{ }_{ }$ & -$^{ }_{ }$ & -$^{ }_{ }$  \\
14 & 515.1 & 537.6 & -$^{ }_{ }$ & -$^{ }_{ }$ & -$^{ }_{ }$ & -$^{ }_{ }$  \\
15 & 577.9 & 541.9 & < 122$^{ }_{ }$ & 10$^{+3}_{-0}$ & 16.5$^{+3}_{-5}$ & 13.5$^{+13}_{-2}$  \\
16 & 529.2 & 542.4 & -$^{ }_{ }$ & -$^{ }_{ }$ & -$^{ }_{ }$ & -$^{ }_{ }$  \\
17 & 625.7 & 546.4 & 13$^{+1}_{-2}$ & -$^{ }_{ }$ & 9$^{+1}_{-1}$ & -$^{ }_{ }$  \\
18 & 544.2 & 547.7 & -$^{ }_{ }$ & -$^{ }_{ }$ & -$^{ }_{ }$ & -$^{ }_{ }$  \\
19 & 479.5 & 549 & -$^{ }_{ }$ & -$^{ }_{ }$ & -$^{ }_{ }$ & -$^{ }_{ }$  \\
20 & 648.7 & 552.2 & 10$^{+3}_{-0}$ & < 313$^{ }_{ }$ & 9.5$^{+1}_{-1}$ & 11.4$^{+1}_{-1}$  \\
21 & 639.5 & 559.6 & < 555$^{ }_{ }$ & 13$^{+1}_{-3}$ & -$^{ }_{ }$ & 12.5$^{+2}_{-1}$  \\
22 & 599.8 & 558.8 & < 43$^{ }_{ }$ & 12$^{+1}_{-2}$ & 19.5$^{+12}_{-2}$ & 28$^{+11}_{-3}$  \\
23 & 622.9 & 568.3 & 100$^{+8}_{-7}$ & 78$^{+10}_{-8}$ & > 32$^{ }_{ }$ & > 32$^{ }_{ }$  \\
24 & 640.9 & 562.9 & 13$^{+1}_{-1}$ & 13$^{+1}_{-3}$ & -$^{ }_{ }$ & 12.5$^{+2}_{-1}$  \\
25 & 616.6 & 577.5 & 78$^{+11}_{-11}$ & 44$^{+7}_{-9}$ & -$^{ }_{ }$ & 11.7$^{+1306}_{-1}$  \\
26 & 574.7 & 582.3 & -$^{ }_{ }$ & 13$^{+1}_{-0}$ & -$^{ }_{ }$ & 14.6$^{+1}_{-1}$  \\
27 & 585.3 & 584.4 & 13$^{+1}_{-1}$ & 32$^{+3}_{-15}$ & > 32$^{ }_{ }$ & > 32$^{ }_{ }$  \\
28 & 529.7 & 588.8 & < 97$^{ }_{ }$ & 13$^{+1}_{-3}$ & -$^{ }_{ }$ & 13.3$^{+1}_{-1}$  \\
29 & 535 & 591.4 & < 70$^{ }_{ }$ & 10$^{+3}_{-0}$ & -$^{ }_{ }$ & > 32$^{ }_{ }$  \\
30 & 482.8 & 600.2 & -$^{ }_{ }$ & 10$^{+3}_{-0}$ & -$^{ }_{ }$ & 10.4$^{+1}_{-1}$  \\
31 & 527.9 & 596.5 & -$^{ }_{ }$ & 73$^{+20}_{-10}$ & -$^{ }_{ }$ & 10.8$^{+1}_{-1}$  \\
32 & 436.1 & 603.5 & -$^{ }_{ }$ & -$^{ }_{ }$ & -$^{ }_{ }$ & -$^{ }_{ }$  \\
33 & 460.5 & 614 & -$^{ }_{ }$ & -$^{ }_{ }$ & -$^{ }_{ }$ & -$^{ }_{ }$  \\
34 & 504.2 & 620 & -$^{ }_{ }$ & 49$^{+20}_{-31}$ & -$^{ }_{ }$ & > 32$^{ }_{ }$  \\
35 & 451.3 & 630.4 & -$^{ }_{ }$ & 66$^{+45}_{-26}$ & -$^{ }_{ }$ & 12.3$^{+1306}_{-1}$  \\
36 & 487.2 & 642.7 & -$^{ }_{ }$ & 574$^{+402}_{-133}$ & -$^{ }_{ }$ & 11.2$^{+1}_{-1}$  \\
37 & 441.6 & 650 & -$^{ }_{ }$ & -$^{ }_{ }$ & -$^{ }_{ }$ & -$^{ }_{ }$  \\
38 & 384.7 & 696.7 & -$^{ }_{ }$ & -$^{ }_{ }$ & -$^{ }_{ }$ & -$^{ }_{ }$  \\
39 & 523.1 & 727 & -$^{ }_{ }$ & -$^{ }_{ }$ & -$^{ }_{ }$ & -$^{ }_{ }$  \\
40 & 470.7 & 736.3 & -$^{ }_{ }$ & -$^{ }_{ }$ & -$^{ }_{ }$ & -$^{ }_{ }$  \\
41 & 528.6 & 757.1 & -$^{ }_{ }$ & -$^{ }_{ }$ & -$^{ }_{ }$ & -$^{ }_{ }$  \\ \hline
    \end{tabularx}
    \label{Table:table_age}
\end{table*}

\subsection{Age estimates}
\label{subsec:ages}
We convert the $W_{CO}$ and $W_{Br\gamma}$ measurements to age estimates using the Starburst99 models shown in Figure \ref{Fig:sb99_CO} (assuming a Salpeter IMF and metallicity of \mbox{Z = 0.020}, see Figures 87b and 101b of \citealt{Leitherer99}). The equivalent width measurements should be approximately independent of extinction, since any reddening has the same multiplicative impact on both the stellar emission and absorption. $W_{Br\gamma}$ decreases monotonically with age, and therefore only one age estimate is associated with each $W_{Br\gamma}$ value. However, $W_{CO}$ oscillates considerably, resulting in a range of possible ages particularly for clusters with \mbox{11 \AA\ $<$ $W_{CO}$ $<$ 13 \AA}. The errors on the age estimates are derived by determining the ages corresponding to the lower and upper boundaries of the 1$\sigma$ confidence interval of the equivalent width values. The fast overall decrease in the expected equivalent width values as a function of age results in relatively small age errors, especially on the estimates derived from $W_{Br\gamma}$ . In cases where multiple ages are consistent with a given  $W_{CO}$ measurement, the final age estimate is the average of all the individual age estimates and the age error encompasses the error intervals of all the individual estimates. We also compare the ages of the clusters detected in CO but not \Brg\ (6/22) to	 the lower age limit implied from the non-detection of \Brg\ (\mbox{$>$ 32 Myr}).

\begin{figure*}
\centerline{\includegraphics[scale=1, clip = true, trim = 0 160 0 0]{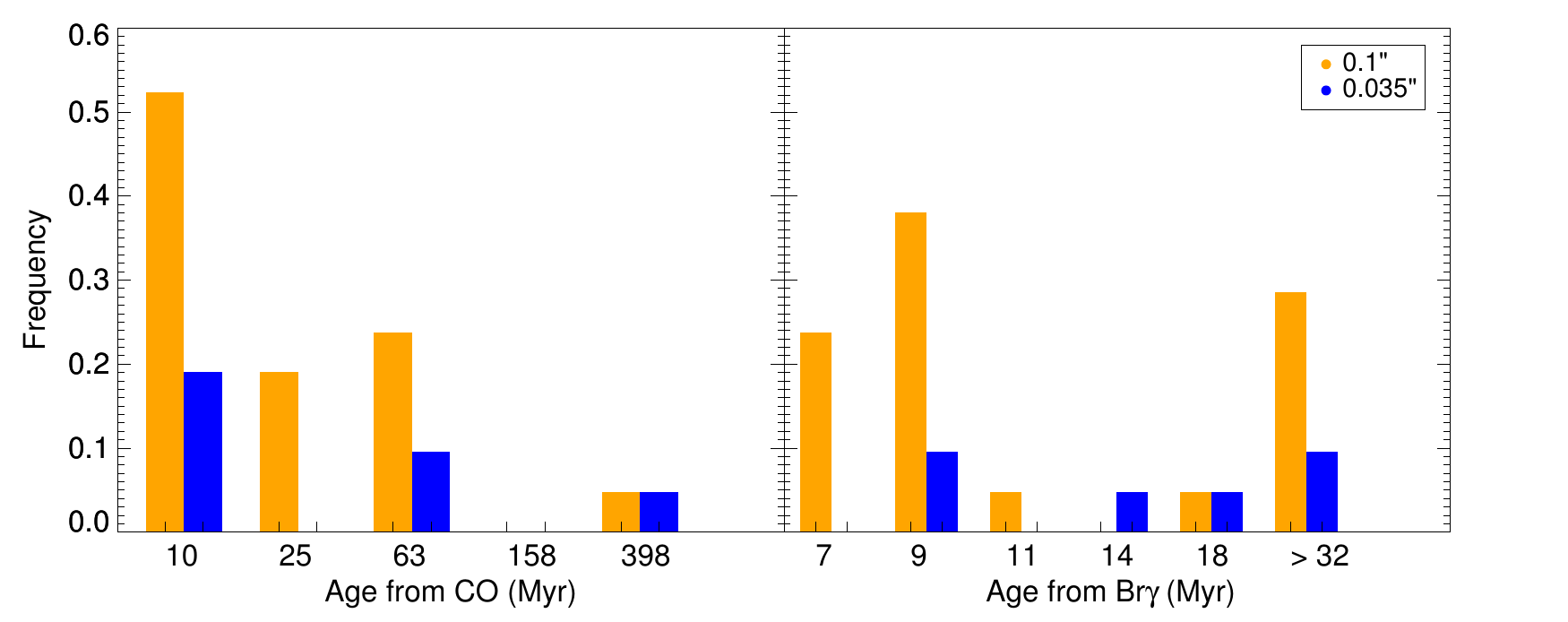}}
\caption{Histograms of star cluster ages derived using (left) $W_{CO}$ and (right) $W_{Br\gamma}$ from (orange) \hs\ and (blue) \ts\ data. The clusters that are detected in CO but not in \Brg\ have ages \mbox{$>$ 32 Myr}. The histogram frequencies are expressed as a fraction of the total spectroscopic cluster sample, such that the orange histograms sum to 1 and the blue histograms sum to 0.52, the fraction of the spectroscopic cluster sample that falls within the \ts\ field. The peak at ages \mbox{$\la$ 20 Myr} in both the $W_{CO}$ and $W_{Br\gamma}$ histograms is indicative of current starburst activity. There is also some evidence for a second peak at \mbox{$\sim$ 65 Myr} in the $W_{CO}$ histograms.}\label{Fig:CO_age}
\end{figure*} 

The age estimates for all clusters lying within the OSIRIS field are listed in Table \ref{Table:table_age} and illustrated in Figure \ref{Fig:age_measurements}. We are able to derive age estimates for 22/25 clusters within the \hs\ field, of which 11 also have age estimates from the \ts\ data. The brown, red, green and blue points in Figure \ref{Fig:age_measurements} indicate age estimates determined using \Brg\ \hs\ data,  \Brg\ \ts\ data, CO \hs\ data and CO \ts\ data respectively. The individual age estimates are consistent with one another within the errors for 68 per cent (15/22) of clusters and within twice the size of the errors for 82 per cent (18/22) of clusters, indicating very good agreement despite the small error bars on many of the measurements. The cluster with the most discrepant age estimates (31) lies within \mbox{90 mas} (\mbox{$<$ 40 pc}) of two young clusters (28 and 29), one or both of which may be responsible for exciting the \Brg\ emission detected at the location of cluster 31. The \Brg\ estimate for cluster 31 is marked with a `+' sign in Figure \ref{Fig:age_measurements} to indicate that it may not be reliable. 

The CO age upper limits are consistent with other age measurements for the same clusters (e.g. from $W_{Br\gamma}$ or $W_{CO}$ in the other plate scale) in 100 per cent (6/6) of cases, and Br$\gamma$ lower age limits are consistent with age estimates for the same clusters in 57 per cent (4/7) of cases. Both of the clusters with discrepant \Brg\ age limits (7 and 36) have neighbouring clusters within \mbox{45 pc}, and therefore the discrepancies may again be attributable to contamination from other clusters. These clusters do not have \ts\ \Brg\ estimates and therefore we are unable to comment on whether the higher angular resolution makes the \ts\ measurements more accurate in these cases. There is no systematic offset between age estimates derived from the same indicator but in different plate scales, indicating that our background subtraction technique is consistent. 

Figure \ref{Fig:CO_age} shows histograms of the age estimates from $W_{CO}$ (left) and $W_{Br\gamma}$ (right), with \ts\ and \hs\ data shown in blue and orange respectively. The histogram frequencies are expressed as a fraction of the total spectroscopic cluster sample, such that if both CO and \Brg\ age estimates were available for all of the clusters in the sample, then the orange histogram would sum to 1 and the blue histogram would sum to 0.52, the fraction of the spectroscopic cluster sample that falls within the \ts\ field. However, the true histogram sums are lower than this due to the limited age range within which each of the age indicators are detectable. There is a peak in all four histograms at ages \mbox{$\la$ 20 Myr}. A total of 70 per cent of the \Brg\ measurements and 54 per cent of the CO measurements fall within this age range, suggesting that MCG08 is currently undergoing a burst of star formation. There is also evidence for a possible second peak at \mbox{$\sim$ 65 Myr}. 29 per cent of CO measurements fall between ages of \mbox{32 - 100 Myr}, which may be evidence for a second burst of star formation in the recent history of the galaxy. The \Brg\ tracer is not sensitive to ages \mbox{$>$ 32 Myr}, so we cannot rely on it to lend evidence for or against such a burst (see Section \ref{subsec:single_double} for further discussion). 

We note that we are unable to combine the individual age estimates to produce a single age estimate for each cluster because we do not have the underlying age probability distribution functions associated with each of the indicators. Instead, we construct `minimal' and `maximal' star cluster age distributions (using the minimum and maximum age estimate respectively for each star cluster) which we compare to models for different star formation histories in Section \ref{subsec:single_double}. 

\section{Recovering the intrinsic star cluster age distribution}
\label{sec:stats}

\subsection{Quantifying selection effects}
\label{subsubsec:quantify_bias}
\label{subsec:biases}
Before analysing the star formation history of MCG08 it is important to confirm that the observed star cluster age distribution is representative of the underlying star cluster population in the nuclear region of the system. Our star cluster sample is comprised of quasi-point sources which are detected at a threshold significance above the background. The luminosity contrast between a star cluster and the background depends on the absolute magnitude of the cluster, the optical depth along the line of sight and the background magnitude. Stellar populations become fainter as they evolve, causing the integrated luminosity of a star cluster to decrease with age. The average background luminosity and optical depth increase towards the centre of MCG08, creating a gradient in the cluster detection threshold between the inner and outer regions of the galaxy.

\begin{table*}
    \caption{Percentage of synthetic clusters identified by our point-source detection algorithm as a function of age and optical depth. `Inner' and `outer' clusters refer to those lying within and outside the OSIRIS \hs\ FOV respectively. The detection fraction of both the inner and outer clusters decreases as age decreases and as optical depth increases. At least 80 per cent of clusters with ages up to 500 (100) Myr are detected at optical depths up to \mbox{$\tau$ = 17} (20). Only 64 percent of 100 Myr old clusters are detected at \mbox{$\tau$ = 22}. The detection fraction drops below 50 per cent for clusters with ages \mbox{$>$ 100 Myr} at an optical depth of \mbox{$\tau$ = 20}.}
    \begin{tabularx}{\textwidth}{RR|RRR|RRR|RRR|RRR|RRR|RRR}
    \hline
    \multicolumn{2}{c}{~} & \multicolumn{3}{c}{$\tau$ = 10} & \multicolumn{3}{c}{$\tau$ = 12} & \multicolumn{3}{c}{$\tau$ = 14} & \multicolumn{3}{c}{$\tau$ = 17} & \multicolumn{3}{c}{$\tau$ = 20} & \multicolumn{3}{c}{$\tau$ = 22} \\ \hline
    \multicolumn{2}{c}{Age (Myr)} & All & Inner & Outer & All & Inner & Outer & All & Inner & Outer & All & Inner & Outer & All & Inner & Outer & All & Inner & Outer \\ \hline
    10 & ~ & 100 & 100 & 100 & 100 & 100 & 100 & 100 & 100 & 100 & 100 & 100 & 100 & 100 & 100 & 100 & 100 & 100 & 100 \\  	
    50 & ~ & 97 & 95 & 100 & 97 & 95 & 100 & 97 & 95 & 100 & 96 & 93 & 100 & 95 & 91 & 100 & 94 & 89 & 100 \\
    65 & ~ & 97 & 95 & 100 & 97 & 95 & 100 & 97 & 95 & 100 & 95 & 91 & 100 & 94 & 89 & 100 & 94 & 89 & 100 \\
    100 & ~ & 97 & 95 & 100 & 96 & 93 & 100 & 95 & 91 & 100 & 94 & 89 & 100 & 92 & 88 & 98 & 64 & 68 & 58 \\
    200 & ~ & 95 & 91 & 100 & 95 & 91 & 100 & 94 & 89 & 100 & 91 & 86 & 98 & 44 & 65 & 16 & 37 & 61 & 5 \\ 
    500 & ~ & 95 & 91 & 100 & 94 & 89 & 100 & 94 & 89 & 100 & 89 & 82 & 98 & 44 & 65 & 16 & 37 & 61 & 5 \\ \hline
	\end{tabularx}
	\label{Table:cluster_detection}
\end{table*}

We quantify the detection limit of our sample by determining the probability that a star cluster will be detected as a function of its age, optical depth and position in the galaxy. We probe a range of ages (10, 50, 65, 100, 200 and \mbox{500 Myr}) and optical depths (\mbox{$\tau$ = 10, 12, 14, 17, 20, 22}) designed to reflect the variety of properties observed within MCG08. The average optical depths (calculated using the \Brg/\Brd\ ratio, assuming Case B recombination in a nebula at \mbox{10,000-20,000 K} with an electron density of \mbox{100 cm$^{-3}$}) within the FOV of the OSIRIS \ts\ and \hs\ data respectively are \mbox{$\tau$ = 14} and \mbox{$\tau$ = 12}, and 86 per cent of the spaxels in the \hs\ FOV for which optical depths could be calculated are consistent with having \mbox{$\tau$ = 17} or less. It is more difficult to place a constraint on the optical depth outside the OSIRIS FOV where no spectroscopic information is available. The SED fitting described briefly in Section \ref{subsec:SEDs} indicates that models with optical depths less than 10 are inconsistent with the derived limits on the cluster F814W and F435W magnitudes.

We take the observed Kp band image of MCG08 and randomly insert 100 synthetic clusters of a single age and optical depth at different locations within 150 pixels (1.5 arcsec) of the nucleus. We require the $x$ and $y$ coordinates of each cluster centroid to be at least 20 pixels from neighbouring clusters to prevent source confusion. We use FSPS models \citep{Conroy10} to extract the Kp band absolute magnitude for a star cluster of the relevant age and optical depth, convert the absolute magnitude to an apparent magnitude using the known redshift of MCG08, and then to counts by applying a reverse flux calibration. We generate the PSF of the synthetic star cluster by re-normalizing the PSF of the tip-tilt star to contain the same total number of counts as expected from the star cluster. As expected, the contrast between the star clusters and the background decreases significantly as the cluster age decreases and as the optical depth increases. 

We apply the FIND algorithm to each of the twenty five images, using the same parameters applied to the original Kp band image in Section \ref{subsec:identification}. We calculate the detection probabilities of clusters internal and external to the OSIRIS \hs\ FOV separately (`inner' and `outer' clusters respectively) to account for the significantly higher background level and optical depth in the inner region of the galaxy. The percentage of total, inner and outer synthetic clusters identified by the cluster detection algorithm for each of the images are listed in Table \ref{Table:cluster_detection}. 

We find that 100 per cent of the \mbox{10 Myr} old clusters can be detected at all considered optical depths and all radii, indicating that our cluster sample is complete for the youngest clusters. At least 90 (85) per cent of the \mbox{50 Myr} (\mbox{100 Myr}) old clusters are detected at all optical depths except \mbox{$\tau$ = 22} where only 64 percent of 100 Myr old clusters are detected. At least 80 per cent of the \mbox{200 Myr} and \mbox{500 Myr} old clusters are detected at optical depths up to and including \mbox{$\tau$ = 17}, but the overall detection rates drop to 44 per cent at \mbox{$\tau$ = 20}. Our analysis indicates that our spectroscopic star cluster sample is \mbox{$\ga$ 80} per cent complete for star clusters with ages $\leq$ 500 Myr over $\sim$86 per cent of the region covered by our observations, and \mbox{$\ga$ 89} per cent complete for star clusters with ages $\leq$ 65 Myr over $\sim$93 per cent of the region covered by our observations. The scarcity of detected star clusters with ages greater than \mbox{100 Myr} is therefore likely to be intrinsic to MCG08 rather than a selection effect.

\begin{figure*}
\centerline{\includegraphics[scale=0.9]{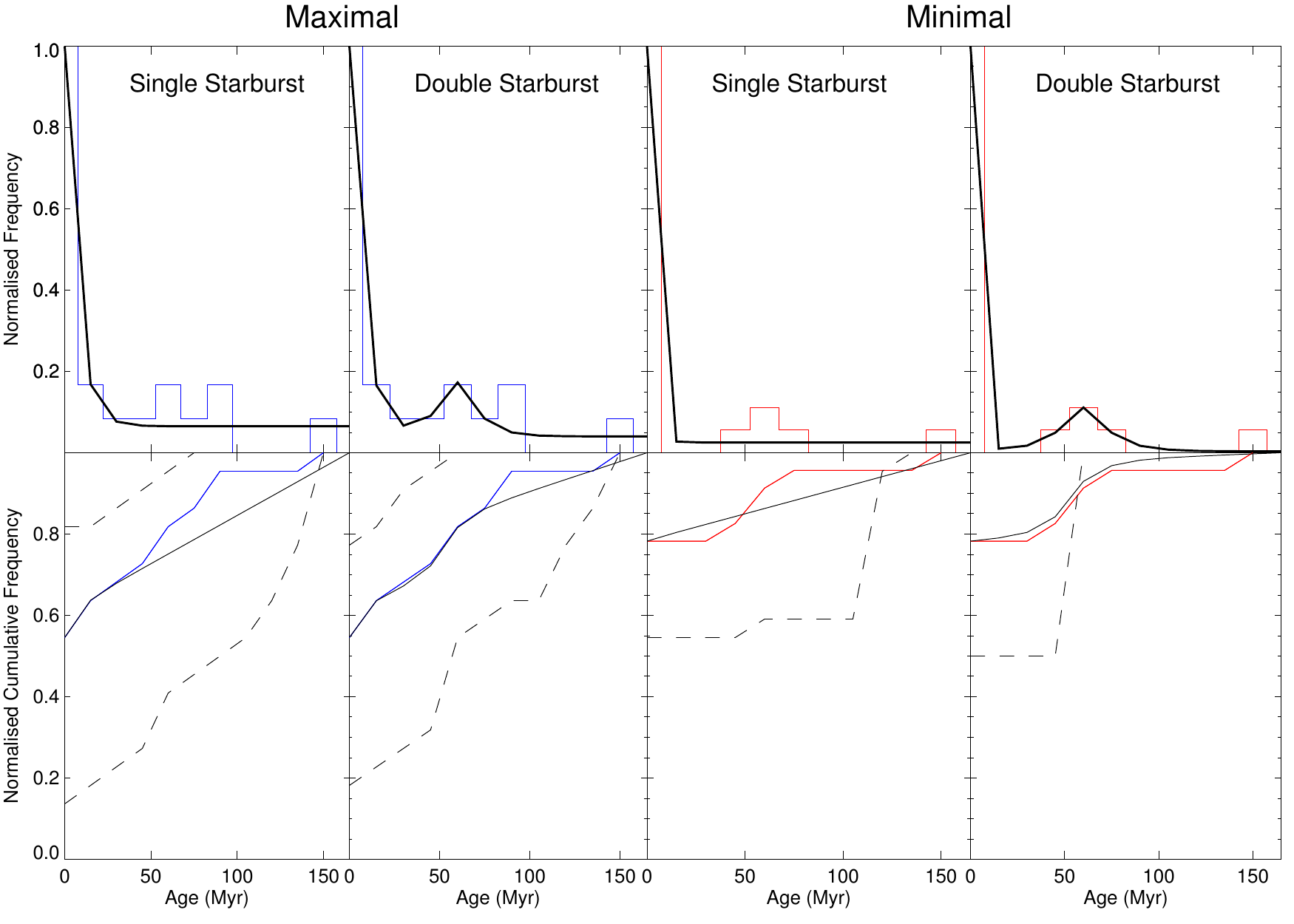}}
\caption{Top panels: (Coloured) normalized maximal and minimal star cluster age distributions and (black) best fit models for the single and double starburst scenarios. Bottom panels: normalized cumulative distribution functions for (coloured) the maximal and minimal age distributions and (solid black) the best fit models. The black dashed curves enclose all 1000 cumulative distribution functions derived from Monte Carlo sampling of the best fit models.}\label{Fig:single_burst_sfh}
\end{figure*} 

\subsection{Single or double starburst?}
\label{subsec:single_double}

\subsubsection{Modelling the underlying star cluster age distribution}
\label{subsec:modelling}
The star cluster age histograms presented in Section \ref{subsec:ages} reveal a clear excess of clusters with ages \mbox{$<$ 20 Myr}, and a possible second peak at \mbox{$\sim$65 Myr}. Although the first peak is very prominent and indicative of a current starburst in the system, the second peak contains a smaller number of clusters and may be consistent with the underlying stochastic continuous star formation history of MCG08. In this section, we construct model star cluster age PDFs for single and double burst star formation histories and compare them to the observed age distributions to determine whether the second burst is statistically significant.

We create minimal and maximal star cluster age distributions using the minimum and maximum age estimates respectively for each of the star clusters. The histograms of these distributions are shown in the top panel of Figure \ref{Fig:single_burst_sfh} (red and blue histograms trace the minimal and maximal distributions respectively). We do not include the maximal age estimate for one star cluster (\mbox{574 Myr}) which is a significant outlier from the rest of the sample. 

The model age PDFs are parametrised as either single or double starburst events (each parametrised as a two-sided exponential decay) superimposed over an underlying continuous star formation history (parametrised as a uniform distribution). The locations of the starburst(s) are fixed, but the e-folding times and peak intensities of the exponentials are left as free parameters. The cumulative distribution function (CDF) of the uniform distribution is not required to sum to one, and the probability scaling is left as a free parameter. The single and double burst models therefore have three and five free parameters respectively.

We use the IDL Levenberg-Marquardt least-squares fitting function MPFIT \citep{Markwardt09} to select the parameters which produce the best fit to each of the observed age PDFs. We assume that the peak of the first and second starburst events occur at the first and second peaks in the observed age PDF (\mbox{7.5 Myr} and \mbox{65 Myr} ago respectively). If the current starburst has not yet reached its peak, then the derived e-folding time of the first burst may not reflect its true e-folding time.

The top panels of Figure \ref{Fig:single_burst_sfh} show the measured maximal (left, blue) and minimal (right, red) age distributions with the best fit models over-plotted in black, convolved to match the binning resolution of the data. The sharp decrease in the number of clusters between the first and second bins of the observed age distributions are reproduced well by the starburst models. The best fit single burst models for the maximal and minimal age distributions imply starburst e-folding times of \mbox{6.8 Myr} and \mbox{2.5 Myr} respectively. The lack of star clusters between the first and second peaks in the minimal age distribution implies a much faster decay of the starburst, thus producing a factor of three smaller e-folding time than implied by the maximal age distribution. 

The minimal age distribution has a clear second peak which is well reproduced by the double burst model. However, the larger age spread of intermediate age (\mbox{50 - 100} Myr) star clusters in the maximal age distribution makes the presence of a second peak unclear. The best fit double burst models for the maximal and minimal age distributions imply starburst e-folding times of [\mbox{7.5 Myr}, \mbox{10 Myr}] and [\mbox{2.5 Myr}, \mbox{12.5 Myr}] respectively. Both fits imply that the second starburst is wider than the first, although this result is not significant given that the difference is less than the time resolution of our age PDFs.

The best-fit continuous star-formation level is significantly larger for the maximal age distribution than the minimal age distribution (in both the single and double burst models) primarily due to the presence of clusters lying in between the two histogram peaks. The continuous star formation level decreases in the double burst model compared to the single burst model, especially for the minimal age distribution in which the vast majority of clusters are contained within either the first or second burst. 

\subsubsection{Statistical tests}
\label{subsubsec:stats}
We use the Kolmogorov-Smirnov (KS) test to determine the probability that our observed star-cluster age distributions are drawn from the best fit single and double starburst age PDF models. The KS statistic gives the maximum difference between the CDFs of the data and the model, and is a good probe of the quality of the match between the data and the model when the data is a good representation of the underlying distribution from which it is drawn. When the sample size is small, stochasticity becomes an important factor in shaping the observed distribution. 

Our sample of 22 star clusters is likely to be too small to effectively sample the underlying age PDF. To reduce the impact of stochastic sampling on our probability calculations, we can apply the two-sided KS test which determines the probability that two samples are drawn from the same parent distribution. We construct mock data samples from each analytical model by randomly sampling 22 data points from a high time resolution discrete distribution (\mbox{$\Delta$ t = 10 kyr}). This is repeated 1000 times to produce 1000 estimates of the two-sided KS statistic. The bottom panels of Figure \ref{Fig:single_burst_sfh} show cumulative distribution functions for the maximal (left, blue) and minimal (right, red) age distributions, as well as the best fit model (solid black line). The dashed curves indicate the range of cumulative distribution functions derived from our Monte Carlo sampling. It is interesting to note that although the cumulative distribution function of the maximal age distribution falls well within the regions covered by the Monte Carlo sampling for both the best fit single and double starburst models, the cumulative distribution of the minimal age distribution is not so consistent with the Monte Carlo samples which are unable to reproduce the star cluster at \mbox{$\sim$150 Myr}. This may suggest that the maximal age distribution is more reflective of the intrinsic star cluster age distribution than the minimal age distribution.

We use the IDL routine \textsc{prob\textunderscore KS} to convert the derived KS statistics into probabilities that the two tested samples are drawn from the same parent distribution. We then calculate the average probability that each observed age distribution is drawn from its best fit single or double burst model (with error given by the standard deviation).

The probabilities that the maximal star cluster age distribution is drawn from a single or double burst star formation history are 92 and 94 per cent respectively, whereas for the minimal age distribution the probabilities are 98 and 99.6 per cent respectively. The standard error of the mean is less than 1 per cent due to the large number of samples used to calculate the probabilities. Based on these numbers alone, the double burst model appears to be (marginally) favoured over the single burst model for both the maximal and minimal age distributions (based on the size of the residuals alone). However, a pure comparison of the probabilities of each of the models does not allow us to discern which one is a better statistical representation of each of the datasets. Increasing the complexity of a model (i.e. by adding an extra starburst) allows more features in the data to be accounted for, but also increases the risk of fitting noise. We determine which model provides a better statistical representation of each of the datasets by calculating the likelihood ratio test statistic $D$:

\begin{align*}
D = - 2 \ln \left(\frac{P_{null}}{P_{alternate}}\right)
\end{align*}

The likelihood ratio compares the probability of the `null' and `alternate' hypotheses (the former of which must be a special case of the latter). In this case, the null hypothesis (single burst model) is derived by setting the scaling of the second burst to zero. The probability distribution of the test statistic can be treated as a $\chi^2$ distribution with degrees of freedom given by the difference in the number of free parameters in the null and alternative hypotheses (in this case, 2). The calculated likelihood ratio statistics for the maximal and minimal age distributions are 0.04 and 0.03 respectively, corresponding to p-values of 0.98. The p-values indicate that the null hypothesis cannot be rejected on the basis of this data alone. 

We note that the results of our statistical analysis are fundamentally limited by our sample size and our conclusions do not change even after applying a completeness correction to our star cluster age distributions. We multiply the frequency of clusters with ages between 25 and 82.5 Myr by 1/0.94, the frequency of clusters with ages between 82.5 and 150 Myr by 1/0.64 and the frequency of clusters with ages $>$150 Myr by 1/0.37 (using the completeness fractions listed for \mbox{$\tau$ = 22} in Table \ref{Table:cluster_detection}). Using these corrected dstributions, we find the probabilities that the maximal star cluster age distribution is drawn from a single or double burst star formation history are 70.5 and 81.4 per cent respectively, whereas for the minimal age distribution the probabilities are 92 and 99 per cent respectively. The likelihood ratios determined for the maximal and minimal age distributions are 0.075 and 0.206 respectively, corresponding to p values of 0.96 and 0.9. Age estimates for a larger sample of star clusters would likely produce larger differences between the probabilities for the two models and therefore provide more definitive results.

\section{Summary and Conclusions}
\label{sec:conclusions}
We have used Keck NIRC2 and OSIRIS to undertake a census of the dust obscured nuclear star cluster population of MCG+08-11-002 (MCG08). With the aid of the Keck LGS-AO system we have obtained high resolution Kp band imaging (\mbox{FWHM $\sim$ 25 pc}) which allows us to resolve 41 star clusters. 25 (13) of these clusters are also covered by our OSIRIS \hs\ (\ts) NIR integral field spectroscopy. We estimate the ages of each of the clusters in our spectroscopic sample using the equivalent widths of the CO 2.3$\mu$m absorption feature and the \Brg\ emission line, which are sensitive to star clusters with ages of \mbox{$\sim$ 10 Myr - 1 Gyr} and \mbox{$\la$ 35 Myr} respectively. We remove the contribution of background sources to the continuum emission of the galaxy by constructing an average `background' spectrum which is then subtracted from every spaxel in the integral field data cube. The CO and \Brg\ equivalent width measurements for each cluster are converted to age estimates using the Starburst99 models of \citet{Leitherer99}. The individual age estimates for each cluster are consistent with one another within the errors for 68 per cent of clusters and within twice the size of the errors for 82 per cent of clusters. There is no systematic offset between age estimates derived from the same indicator but in different plate scales, indicating that our background subtraction technique is consistent. 

The star cluster age distribution of MCG08 has at least one clear peak. 70 per cent of the \Brg\ age measurements and 54 per cent of the CO measurements (averaged across both the \ts\ and the \hs\ data) fall within \mbox{0 - 20 Myr}. There is also some evidence for a second peak in the age distribution at \mbox{$\sim$65 Myr}, with 29 per cent of the CO measurements falling between ages of 32 and \mbox{100 Myr}. Our analysis in Section \ref{subsec:biases} indicates that our star cluster sample is \mbox{$\ga$ 80} per cent complete for star clusters with ages $\leq$ 500 Myr over $\sim$86 per cent of the region covered by our observations, and \mbox{$\ga$ 89} per cent complete for star clusters with ages $\leq$ 65 Myr over $\sim$93 per cent of the region covered by our observations. 

We investigate whether the observed star cluster age distribution of MCG08 is more consistent with a single or double starburst star formation history by fitting model star cluster age distributions to the observed distribution. Without the underlying probability distribution functions associated with each of the age indicators, we are unable to combine the age estimates to determine the maximum likelihood age of each cluster. Instead, we fit models to the minimal and maximal star cluster age distributions, constructed using the minimum and maximum age estimate for each star cluster respectively. We account for stochastic sampling of the underlying age distribution by randomly sampling 22 data points from each model to create `mock data samples'. The probability that the observed distributions are drawn from each of the models is given by the average of the probabilities derived from the two-sided KS test between each of the 1000 mock data samples and the data. We find that there is a greater than 90 per cent chance that the observed age distribution is drawn from either the single or double burst model models, but the likelihood ratio test indicates that our sample size is not sufficient to discriminate between the two models. This conclusion does not change even after applying completeness corrections to our age distributions.

Galaxy merger simulations predict that the star formation histories of merging systems should approximately trace the separation of the progenitor systems and can therefore provide vital insights into their merger timelines. The optical and NIR images of MCG08 indicate that it is a single system and has reached final coalescence. The starburst event responsible for producing the excess of star clusters with ages \mbox{$<$ 20 Myr} is likely to have been triggered by rapid gas inflow during this coalescence. If further data confirm that a second burst did occur in the nuclear region of MCG08, then it was likely associated with the previous close passage of the galaxy nuclei (either as part of the coalescence event or the final pericentre passage preceding coalescence). If, however, the second burst is found to be statistically insignificant, a number of scenarios are possible. The system may have had a long merger timescale and the star clusters from the previous close passage are too old and faint to be detected. Alternatively, the morphologies and initial orbital parameters of the progenitor systems may be such that there was no significant starburst associated with the previous passage \citep[e.g.][]{Mihos96, DiMatteo07}. A third possibility is that a second burst did occur in the recent history of the system, but a significant fraction of the star formation occurred outside of the OSIRIS FOV and is therefore not detected \citep[e.g.][]{Hopkins13}. Comparison of deep, wide-field star cluster age distributions with detailed dynamical models of galaxy mergers \citep[see e.g.][]{Privon13} will provide observational insight into the fuelling of merger-driven starbursts as a function of merger stage, galaxy morphology and orbital parameters.

Our results add to an increasing body of research indicating that star cluster age distributions encode the recent merger histories of their host systems and may therefore be important probes of merger stage. NIR integral field spectroscopy is a valuable tool for examining the star cluster populations of heavily obscured systems, providing resolved temporal and spatial insights into star formation and the building of stellar mass during the most rapid periods of galaxy evolution. 

\section{Acknowledgements}
The authors would like to thank Charlie Conroy and Joshua Barnes for insightful conversations which significantly improved the clarity of this manuscript. Based on observations made with the NASA/ESA Hubble Space Telescope, and obtained from the Hubble Legacy Archive, which is a collaboration between the Space Telescope Science Institute (STScI/NASA), the Space Telescope European Coordinating Facility (ST-ECF/ESA) and the Canadian Astronomy Data Centre (CADC/NRC/CSA). Some of the data presented herein were obtained at the W.M. Keck Observatory, which is operated as a scientific partnership among the California Institute of Technology, the University of California and the National Aeronautics and Space Administration. The Observatory was made possible by the generous financial support of the W.M. Keck Foundation. The authors wish to recognize and acknowledge the very significant cultural role and reverence that the summit of Mauna Kea has always had within the indigenous Hawaiian community. We are most fortunate to have the opportunity to conduct observations from this mountain. 

\bibliography{mybib}

\begin{thebibliography}{74}
\expandafter\ifx\csname natexlab\endcsname\relax\def\natexlab#1{#1}\fi

\bibitem[{{Armus}, {Heckman} \& {Miley}(1987){Armus}, {Heckman}, \&
  {Miley}}]{Armus87}
{Armus} L., {Heckman} T., {Miley} G., 1987, \aj, 94, 831

\bibitem[{{Armus} {et~al}\mbox{.}(2009){Armus}, {Mazzarella}, {Evans},
  {Surace}, {Sanders}, {Iwasawa}, {Frayer}, {Howell}, {Chan}, {Petric},
  {Vavilkin}, {Kim}, {Haan}, {Inami}, {Murphy}, {Appleton}, {Barnes}, {Bothun},
  {Bridge}, {Charmandaris}, {Jensen}, {Kewley}, {Lord}, {Madore}, {Marshall},
  {Melbourne}, {Rich}, {Satyapal}, {Schulz}, {Spoon}, {Sturm}, {U}, {Veilleux},
  \& {Xu}}]{Armus09}
{Armus} L. {et~al.}, 2009, \pasp, 121, 559

\bibitem[{{Barnes}(1992)}]{Barnes92}
{Barnes} J.~E., 1992, \apj, 393, 484

\bibitem[{{Barnes} \& {Hibbard}(2009)}]{Barnes09}
{Barnes} J.~E., {Hibbard} J.~E., 2009, \aj, 137, 3071

\bibitem[{{Barton}, {Geller} \& {Kenyon}(2000){Barton}, {Geller}, \&
  {Kenyon}}]{Barton00}
{Barton} E.~J., {Geller} M.~J., {Kenyon} S.~J., 2000, \apj, 530, 660

\bibitem[{{Calzetti}(2001)}]{Calzetti01}
{Calzetti} D., 2001, \pasp, 113, 1449

\bibitem[{{Chien} {et~al}\mbox{.}(2007){Chien}, {Barnes}, {Kewley}, \&
  {Chambers}}]{Chien07}
{Chien} L.-H., {Barnes} J.~E., {Kewley} L.~J., {Chambers} K.~C., 2007, \apjl,
  660, L105

\bibitem[{{Chilingarian} {et~al}\mbox{.}(2010){Chilingarian}, {Di Matteo},
  {Combes}, {Melchior}, \& {Semelin}}]{Chilingarian10}
{Chilingarian} I.~V., {Di Matteo} P., {Combes} F., {Melchior} A.-L., {Semelin}
  B., 2010, \aap, 518, A61

\bibitem[{{Clements} {et~al}\mbox{.}(1996){Clements}, {Sutherland}, {McMahon},
  \& {Saunders}}]{Clements96}
{Clements} D.~L., {Sutherland} W.~J., {McMahon} R.~G., {Saunders} W., 1996,
  \mnras, 279, 477

\bibitem[{{Conroy} \& {Gunn}(2010)}]{Conroy10}
{Conroy} C., {Gunn} J.~E., 2010, \apj, 712, 833

\bibitem[{{Desai} {et~al}\mbox{.}(2007){Desai}, {Armus}, {Spoon},
  {Charmandaris}, {Bernard-Salas}, {Brandl}, {Farrah}, {Soifer}, {Teplitz},
  {Ogle}, {Devost}, {Higdon}, {Marshall}, \& {Houck}}]{Desai07}
{Desai} V. {et~al.}, 2007, \apj, 669, 810

\bibitem[{{Di Matteo} {et~al}\mbox{.}(2007){Di Matteo}, {Combes}, {Melchior},
  \& {Semelin}}]{DiMatteo07}
{Di Matteo} P., {Combes} F., {Melchior} A.-L., {Semelin} B., 2007, \aap, 468,
  61

\bibitem[{{Di Matteo}, {Springel} \& {Hernquist}(2005){Di Matteo}, {Springel},
  \& {Hernquist}}]{DiMatteo05}
{Di Matteo} T., {Springel} V., {Hernquist} L., 2005, \nat, 433, 604

\bibitem[{{D{\'{\i}}az-Santos} {et~al}\mbox{.}(2010){D{\'{\i}}az-Santos},
  {Charmandaris}, {Armus}, {Petric}, {Howell}, {Murphy}, {Mazzarella},
  {Veilleux}, {Bothun}, {Inami}, {Appleton}, {Evans}, {Haan}, {Marshall},
  {Sanders}, {Stierwalt}, \& {Surace}}]{Diaz-Santos10}
{D{\'{\i}}az-Santos} T. {et~al.}, 2010, \apj, 723, 993

\bibitem[{{D{\'{\i}}az-Santos} {et~al}\mbox{.}(2011){D{\'{\i}}az-Santos},
  {Charmandaris}, {Armus}, {Stierwalt}, {Haan}, {Mazzarella}, {Howell},
  {Veilleux}, {Murphy}, {Petric}, {Appleton}, {Evans}, {Sanders}, \&
  {Surace}}]{Diaz-Santos11}
{D{\'{\i}}az-Santos} T. {et~al.}, 2011, \apj, 741, 32

\bibitem[{{Do} {et~al}\mbox{.}(2013){Do}, {Lu}, {Ghez}, {Morris}, {Yelda},
  {Martinez}, {Wright}, \& {Matthews}}]{Do13}
{Do} T., {Lu} J.~R., {Ghez} A.~M., {Morris} M.~R., {Yelda} S., {Martinez}
  G.~D., {Wright} S.~A., {Matthews} K., 2013, \apj, 764, 154

\bibitem[{{Ellison} {et~al}\mbox{.}(2013){Ellison}, {Mendel}, {Scudder},
  {Patton}, \& {Palmer}}]{Ellison13}
{Ellison} S.~L., {Mendel} J.~T., {Scudder} J.~M., {Patton} D.~R., {Palmer}
  M.~J.~D., 2013, \mnras, 430, 3128

\bibitem[{{Ellison} {et~al}\mbox{.}(2008){Ellison}, {Patton}, {Simard}, \&
  {McConnachie}}]{Ellison08}
{Ellison} S.~L., {Patton} D.~R., {Simard} L., {McConnachie} A.~W., 2008, \aj,
  135, 1877

\bibitem[{{Fischera} \& {Dopita}(2005)}]{Fischera05}
{Fischera} J., {Dopita} M., 2005, \apj, 619, 340

\bibitem[{{Freedman Woods} {et~al}\mbox{.}(2010){Freedman Woods}, {Geller},
  {Kurtz}, {Westra}, {Fabricant}, \& {Dell'Antonio}}]{FreedmanWoods10}
{Freedman Woods} D., {Geller} M.~J., {Kurtz} M.~J., {Westra} E., {Fabricant}
  D.~G., {Dell'Antonio} I., 2010, \aj, 139, 1857

\bibitem[{{Gilbert} {et~al}\mbox{.}(2000){Gilbert}, {Graham}, {McLean},
  {Becklin}, {Figer}, {Larkin}, {Levenson}, {Teplitz}, \& {Wilcox}}]{Gilbert00}
{Gilbert} A.~M. {et~al.}, 2000, \apjl, 533, L57

\bibitem[{{Haan} {et~al}\mbox{.}(2011){Haan}, {Surace}, {Armus}, {Evans},
  {Howell}, {Mazzarella}, {Kim}, {Vavilkin}, {Inami}, {Sanders}, {Petric},
  {Bridge}, {Melbourne}, {Charmandaris}, {Diaz-Santos}, {Murphy}, {U},
  {Stierwalt}, \& {Marshall}}]{Haan11}
{Haan} S. {et~al.}, 2011, \aj, 141, 100

\bibitem[{{Hawarden} {et~al}\mbox{.}(2001){Hawarden}, {Leggett}, {Letawsky},
  {Ballantyne}, \& {Casali}}]{Hawarden01}
{Hawarden} T.~G., {Leggett} S.~K., {Letawsky} M.~B., {Ballantyne} D.~R.,
  {Casali} M.~M., 2001, \mnras, 325, 563

\bibitem[{{Hinshaw} {et~al}\mbox{.}(2009){Hinshaw}, {Weiland}, {Hill},
  {Odegard}, {Larson}, {Bennett}, {Dunkley}, {Gold}, {Greason}, {Jarosik},
  {Komatsu}, {Nolta}, {Page}, {Spergel}, {Wollack}, {Halpern}, {Kogut},
  {Limon}, {Meyer}, {Tucker}, \& {Wright}}]{Hinshaw09}
{Hinshaw} G. {et~al.}, 2009, \apjs, 180, 225

\bibitem[{{Homeier}, {Gallagher} \& {Pasquali}(2002){Homeier}, {Gallagher}, \&
  {Pasquali}}]{Homeier02}
{Homeier} N., {Gallagher}, III J.~S., {Pasquali} A., 2002, \aap, 391, 857

\bibitem[{{Hopkins} {et~al}\mbox{.}(2013){Hopkins}, {Cox}, {Hernquist},
  {Narayanan}, {Hayward}, \& {Murray}}]{Hopkins13}
{Hopkins} P.~F., {Cox} T.~J., {Hernquist} L., {Narayanan} D., {Hayward} C.~C.,
  {Murray} N., 2013, \mnras, 430, 1901

\bibitem[{{Howell} {et~al}\mbox{.}(2010){Howell}, {Armus}, {Mazzarella},
  {Evans}, {Surace}, {Sanders}, {Petric}, {Appleton}, {Bothun}, {Bridge},
  {Chan}, {Charmandaris}, {Frayer}, {Haan}, {Inami}, {Kim}, {Lord}, {Madore},
  {Melbourne}, {Schulz}, {U}, {Vavilkin}, {Veilleux}, \& {Xu}}]{Howell10}
{Howell} J.~H. {et~al.}, 2010, \apj, 715, 572

\bibitem[{{Ishida}(2004)}]{Ishida04}
{Ishida} C.~M., 2004, PhD thesis, UNIVERSITY OF HAWAI'I

\bibitem[{{Kartaltepe} {et~al}\mbox{.}(2012){Kartaltepe}, {Dickinson},
  {Alexander}, {Bell}, {Dahlen}, {Elbaz}, {Faber}, {Lotz}, {McIntosh},
  {Wiklind}, {Altieri}, {Aussel}, {Bethermin}, {Bournaud}, {Charmandaris},
  {Conselice}, {Cooray}, {Dannerbauer}, {Dav{\'e}}, {Dunlop}, {Dekel},
  {Ferguson}, {Grogin}, {Hwang}, {Ivison}, {Kocevski}, {Koekemoer}, {Koo},
  {Lai}, {Leiton}, {Lucas}, {Lutz}, {Magdis}, {Magnelli}, {Morrison}, {Mozena},
  {Mullaney}, {Newman}, {Pope}, {Popesso}, {van der Wel}, {Weiner}, \&
  {Wuyts}}]{Kartaltepe12}
{Kartaltepe} J.~S. {et~al.}, 2012, \apj, 757, 23

\bibitem[{{Lambas} {et~al}\mbox{.}(2003){Lambas}, {Tissera}, {Alonso}, \&
  {Coldwell}}]{Lambas03}
{Lambas} D.~G., {Tissera} P.~B., {Alonso} M.~S., {Coldwell} G., 2003, \mnras,
  346, 1189

\bibitem[{{Larkin} {et~al}\mbox{.}(2006){Larkin}, {Barczys}, {Krabbe},
  {Adkins}, {Aliado}, {Amico}, {Brims}, {Campbell}, {Canfield}, {Gasaway},
  {Honey}, {Iserlohe}, {Johnson}, {Kress}, {LaFreniere}, {Lyke}, {Magnone},
  {Magnone}, {McElwain}, {Moon}, {Quirrenbach}, {Skulason}, {Song}, {Spencer},
  {Weiss}, \& {Wright}}]{Larkin06}
{Larkin} J. {et~al.}, 2006, in Society of Photo-Optical Instrumentation
  Engineers (SPIE) Conference Series, Vol. 6269, Society of Photo-Optical
  Instrumentation Engineers (SPIE) Conference Series

\bibitem[{{Leitherer} {et~al}\mbox{.}(1999){Leitherer}, {Schaerer}, {Goldader},
  {Delgado}, {Robert}, {Kune}, {de Mello}, {Devost}, \&
  {Heckman}}]{Leitherer99}
{Leitherer} C. {et~al.}, 1999, \apjs, 123, 3

\bibitem[{{Lejeune}, {Cuisinier} \& {Buser}(1997){Lejeune}, {Cuisinier}, \&
  {Buser}}]{Lejeune97}
{Lejeune} T., {Cuisinier} F., {Buser} R., 1997, \aaps, 125, 229

\bibitem[{{Lejeune}, {Cuisinier} \& {Buser}(1998){Lejeune}, {Cuisinier}, \&
  {Buser}}]{Lejeune98}
{Lejeune} T., {Cuisinier} F., {Buser} R., 1998, \aaps, 130, 65

\bibitem[{{Lotz} {et~al}\mbox{.}(2008){Lotz}, {Jonsson}, {Cox}, \&
  {Primack}}]{Lotz08}
{Lotz} J.~M., {Jonsson} P., {Cox} T.~J., {Primack} J.~R., 2008, \mnras, 391,
  1137

\bibitem[{{Lu}(2008)}]{Lu08}
{Lu} J.~R., 2008, PhD thesis, University of California, Los Angeles

\bibitem[{{Marigo} \& {Girardi}(2007)}]{Marigo07}
{Marigo} P., {Girardi} L., 2007, \aap, 469, 239

\bibitem[{{Marigo} {et~al}\mbox{.}(2008){Marigo}, {Girardi}, {Bressan},
  {Groenewegen}, {Silva}, \& {Granato}}]{Marigo08}
{Marigo} P., {Girardi} L., {Bressan} A., {Groenewegen} M.~A.~T., {Silva} L.,
  {Granato} G.~L., 2008, \aap, 482, 883

\bibitem[{{Markwardt}(2009)}]{Markwardt09}
{Markwardt} C.~B., 2009, in Astronomical Society of the Pacific Conference
  Series, Vol. 411, Astronomical Data Analysis Software and Systems XVIII,
  {Bohlender} D.~A., {Durand} D., {Dowler} P., eds., p. 251

\bibitem[{{Max} {et~al}\mbox{.}(2005){Max}, {Canalizo}, {Macintosh}, {Raschke},
  {Whysong}, {Antonucci}, \& {Schneider}}]{Max05}
{Max} C.~E., {Canalizo} G., {Macintosh} B.~A., {Raschke} L., {Whysong} D.,
  {Antonucci} R., {Schneider} G., 2005, \apj, 621, 738

\bibitem[{{Medling} {et~al}\mbox{.}(2014){Medling}, {U}, {Guedes}, {Max},
  {Mayer}, {Armus}, {Holden}, {Ro{\v s}kar}, \& {Sanders}}]{Medling14}
{Medling} A.~M. {et~al.}, 2014, \apj, 784, 70

\bibitem[{{Medling} {et~al}\mbox{.}(2015){Medling}, {U}, {Max}, {Sanders},
  {Armus}, {Holden}, {Mieda}, {Wright}, \& {Larkin}}]{Medling15}
{Medling} A.~M. {et~al.}, 2015, \apj, 803, 61

\bibitem[{{Melnick} \& {Mirabel}(1990)}]{Melnick90}
{Melnick} J., {Mirabel} I.~F., 1990, \aap, 231, L19

\bibitem[{{Mihos} \& {Hernquist}(1994)}]{Mihos94}
{Mihos} J.~C., {Hernquist} L., 1994, \apjl, 431, L9

\bibitem[{{Mihos} \& {Hernquist}(1996)}]{Mihos96}
{Mihos} J.~C., {Hernquist} L., 1996, \apj, 464, 641

\bibitem[{{Miller} {et~al}\mbox{.}(1997){Miller}, {Whitmore}, {Schweizer}, \&
  {Fall}}]{Miller97}
{Miller} B.~W., {Whitmore} B.~C., {Schweizer} F., {Fall} S.~M., 1997, \aj, 114,
  2381

\bibitem[{{Mulia}, {Chandar} \& {Whitmore}(2015){Mulia}, {Chandar}, \&
  {Whitmore}}]{Mulia15}
{Mulia} A.~J., {Chandar} R., {Whitmore} B.~C., 2015, \apj, 805, 99

\bibitem[{{Origlia}, {Moorwood} \& {Oliva}(1993){Origlia}, {Moorwood}, \&
  {Oliva}}]{Origlia93}
{Origlia} L., {Moorwood} A.~F.~M., {Oliva} E., 1993, \aap, 280, 536

\bibitem[{{Osterbrock} \& {Ferland}(2006)}]{Osterbrock06}
{Osterbrock} D.~E., {Ferland} G.~J., 2006, {Astrophysics of gaseous nebulae and
  active galactic nuclei}

\bibitem[{{Pasquali}, {de Grijs} \& {Gallagher}(2003){Pasquali}, {de Grijs}, \&
  {Gallagher}}]{Pasquali03}
{Pasquali} A., {de Grijs} R., {Gallagher} J.~S., 2003, \mnras, 345, 161

\bibitem[{{Patton} {et~al}\mbox{.}(2013){Patton}, {Torrey}, {Ellison},
  {Mendel}, \& {Scudder}}]{Patton13}
{Patton} D.~R., {Torrey} P., {Ellison} S.~L., {Mendel} J.~T., {Scudder} J.~M.,
  2013, \mnras, 433, L59

\bibitem[{{Petric} {et~al}\mbox{.}(2011){Petric}, {Armus}, {Howell}, {Chan},
  {Mazzarella}, {Evans}, {Surace}, {Sanders}, {Appleton}, {Charmandaris},
  {D{\'{\i}}az-Santos}, {Frayer}, {Haan}, {Inami}, {Iwasawa}, {Kim}, {Madore},
  {Marshall}, {Spoon}, {Stierwalt}, {Sturm}, {U}, {Vavilkin}, \&
  {Veilleux}}]{Petric11}
{Petric} A.~O. {et~al.}, 2011, \apj, 730, 28

\bibitem[{{Pollack}, {Max} \& {Schneider}(2007){Pollack}, {Max}, \&
  {Schneider}}]{Pollack07}
{Pollack} L.~K., {Max} C.~E., {Schneider} G., 2007, \apj, 660, 288

\bibitem[{{Privon} {et~al}\mbox{.}(2013){Privon}, {Barnes}, {Evans}, {Hibbard},
  {Yun}, {Mazzarella}, {Armus}, \& {Surace}}]{Privon13}
{Privon} G.~C., {Barnes} J.~E., {Evans} A.~S., {Hibbard} J.~E., {Yun} M.~S.,
  {Mazzarella} J.~M., {Armus} L., {Surace} J., 2013, \apj, 771, 120

\bibitem[{{Renaud} {et~al}\mbox{.}(2014){Renaud}, {Bournaud}, {Kraljic}, \&
  {Duc}}]{Renaud14}
{Renaud} F., {Bournaud} F., {Kraljic} K., {Duc} P.-A., 2014, \mnras, 442, L33

\bibitem[{{Sanders} \& {Mirabel}(1996)}]{Sanders96}
{Sanders} D.~B., {Mirabel} I.~F., 1996, \araa, 34, 749

\bibitem[{{Sanders} {et~al}\mbox{.}(1988){Sanders}, {Soifer}, {Elias},
  {Madore}, {Matthews}, {Neugebauer}, \& {Scoville}}]{Sanders88}
{Sanders} D.~B., {Soifer} B.~T., {Elias} J.~H., {Madore} B.~F., {Matthews} K.,
  {Neugebauer} G., {Scoville} N.~Z., 1988, \apj, 325, 74

\bibitem[{{Schawinski} {et~al}\mbox{.}(2010){Schawinski}, {Evans}, {Virani},
  {Urry}, {Keel}, {Natarajan}, {Lintott}, {Manning}, {Coppi}, {Kaviraj},
  {Bamford}, {J{\'o}zsa}, {Garrett}, {van Arkel}, {Gay}, \&
  {Fortson}}]{Schawinski10}
{Schawinski} K. {et~al.}, 2010, \apjl, 724, L30

\bibitem[{{Scudder} {et~al}\mbox{.}(2012){Scudder}, {Ellison}, {Torrey},
  {Patton}, \& {Mendel}}]{Scudder12}
{Scudder} J.~M., {Ellison} S.~L., {Torrey} P., {Patton} D.~R., {Mendel} J.~T.,
  2012, \mnras, 426, 549

\bibitem[{{Snyder} {et~al}\mbox{.}(2015){Snyder}, {Lotz}, {Moody}, {Peth},
  {Freeman}, {Ceverino}, {Primack}, \& {Dekel}}]{Snyder15}
{Snyder} G.~F., {Lotz} J., {Moody} C., {Peth} M., {Freeman} P., {Ceverino} D.,
  {Primack} J., {Dekel} A., 2015, \mnras, 451, 4290

\bibitem[{{U} {et~al}\mbox{.}(2013){U}, {Medling}, {Sanders}, {Max}, {Armus},
  {Iwasawa}, {Evans}, {Kewley}, \& {Fazio}}]{U13}
{U} V. {et~al.}, 2013, \apj, 775, 115

\bibitem[{{van Dam} {et~al}\mbox{.}(2006){van Dam}, {Bouchez}, {Le Mignant},
  {Johansson}, {Wizinowich}, {Campbell}, {Chin}, {Hartman}, {Lafon}, {Stomski},
  \& {Summers}}]{vanDam06}
{van Dam} M.~A. {et~al.}, 2006, \pasp, 118, 310

\bibitem[{{van Dam}, {Le Mignant} \& {Macintosh}(2004){van Dam}, {Le Mignant},
  \& {Macintosh}}]{vanDam04}
{van Dam} M.~A., {Le Mignant} D., {Macintosh} B.~A., 2004, \ao, 43, 5458

\bibitem[{{Veilleux}, {Kim} \& {Sanders}(2002){Veilleux}, {Kim}, \&
  {Sanders}}]{Veilleux02}
{Veilleux} S., {Kim} D.-C., {Sanders} D.~B., 2002, \apjs, 143, 315

\bibitem[{{Veilleux} {et~al}\mbox{.}(1995){Veilleux}, {Kim}, {Sanders},
  {Mazzarella}, \& {Soifer}}]{Veilleux95}
{Veilleux} S., {Kim} D.-C., {Sanders} D.~B., {Mazzarella} J.~M., {Soifer}
  B.~T., 1995, \apjs, 98, 171

\bibitem[{{Vogt}, {Dopita} \& {Kewley}(2013){Vogt}, {Dopita}, \&
  {Kewley}}]{Vogt13}
{Vogt} F.~P.~A., {Dopita} M.~A., {Kewley} L.~J., 2013, \apj, 768, 151

\bibitem[{{Westera} {et~al}\mbox{.}(2002){Westera}, {Lejeune}, {Buser},
  {Cuisinier}, \& {Bruzual}}]{Westera02}
{Westera} P., {Lejeune} T., {Buser} R., {Cuisinier} F., {Bruzual} G., 2002,
  \aap, 381, 524

\bibitem[{{Whitmore} \& {Zhang}(2002)}]{Whitmore02}
{Whitmore} B.~C., {Zhang} Q., 2002, \aj, 124, 1418

\bibitem[{{Whitmore} {et~al}\mbox{.}(1999){Whitmore}, {Zhang}, {Leitherer},
  {Fall}, {Schweizer}, \& {Miller}}]{Whitmore99}
{Whitmore} B.~C., {Zhang} Q., {Leitherer} C., {Fall} S.~M., {Schweizer} F.,
  {Miller} B.~W., 1999, \aj, 118, 1551

\bibitem[{{Wilson} {et~al}\mbox{.}(2006){Wilson}, {Harris}, {Longden}, \&
  {Scoville}}]{Wilson06}
{Wilson} C.~D., {Harris} W.~E., {Longden} R., {Scoville} N.~Z., 2006, \apj,
  641, 763

\bibitem[{{Wizinowich} {et~al}\mbox{.}(2000){Wizinowich}, {Acton}, {Shelton},
  {Stomski}, {Gathright}, {Ho}, {Lupton}, {Tsubota}, {Lai}, {Max}, {Brase},
  {An}, {Avicola}, {Olivier}, {Gavel}, {Macintosh}, {Ghez}, \&
  {Larkin}}]{Wizinowich00}
{Wizinowich} P. {et~al.}, 2000, \pasp, 112, 315

\bibitem[{{Wizinowich} {et~al}\mbox{.}(2006){Wizinowich}, {Le Mignant},
  {Bouchez}, {Campbell}, {Chin}, {Contos}, {van Dam}, {Hartman}, {Johansson},
  {Lafon}, {Lewis}, {Stomski}, {Summers}, {Brown}, {Danforth}, {Max}, \&
  {Pennington}}]{Wizinowich06}
{Wizinowich} P.~L. {et~al.}, 2006, \pasp, 118, 297

\bibitem[{{Yelda} {et~al}\mbox{.}(2010){Yelda}, {Lu}, {Ghez}, {Clarkson},
  {Anderson}, {Do}, \& {Matthews}}]{Yelda10}
{Yelda} S., {Lu} J.~R., {Ghez} A.~M., {Clarkson} W., {Anderson} J., {Do} T.,
  {Matthews} K., 2010, \apj, 725, 331

\bibitem[{{Yuan}, {Kewley} \& {Sanders}(2010){Yuan}, {Kewley}, \&
  {Sanders}}]{Yuan10}
{Yuan} T.-T., {Kewley} L.~J., {Sanders} D.~B., 2010, \apj, 709, 884

\end{thebibliography}

\newpage
\appendix

\section{PSF photometry}
\label{sec:appendix}

We calculate the magnitudes of each of the 41 Kp band detected star clusters in the NIRC2 Kp and J and HST/ACS F814W and F435W bands using PSF photometry (as summarised in Section \ref{subsec:SEDs}). The majority of the derived F814W and F435W fluxes are strict upper limits due to the presence of obscuring dust which prevents the star clusters from being detectable as point-like sources at visible wavelengths. We calculate SED model grids using the Flexible Stellar Population Synthesis (FSPS) code \citep{Conroy10}, and compare the \mbox{J-Kp} color of each cluster with each model in the grids to derive age probability distribution functions (PDFs). Unfortunately our F814W and F435W magnitude limits are not sufficient to break the age-optical depth degeneracy and therefore the ages of the star clusters remain unconstrained. We include a full description of our photometric analysis in this Appendix for completeness, and as a reference in the event that deeper optical images of this galaxy become available in the future.

\begin{figure*}
\centerline{\includegraphics[scale=1,clip=true,trim = 0 210 0 0]{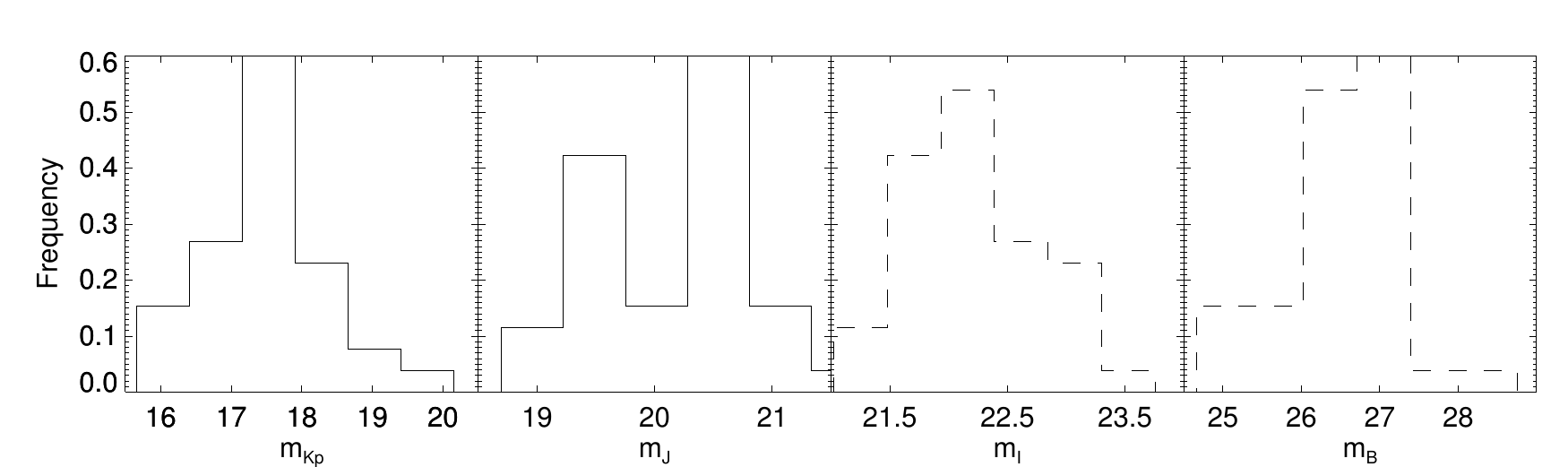}}
\caption{Histograms of the Kp, J, F814W and F435W band apparent magnitudes of our star cluster sample. The majority of the F814W and F435W magnitudes are limits only and therefore the intrinsic magnitude distributions are largely unconstrained.}\label{Fig:maghist}
\end{figure*}

\subsection{PSF characterisation}
\label{subsec:psf_phot}
We calculate the magnitudes of each of the 41 star clusters in all four photometric bands using PSF photometry. The high space density of star clusters in the nuclear region of MCG08 coupled with the diffuse galactic background emission makes it difficult to construct apertures containing the majority of the star cluster light without introducing significant contamination from other sources. Instead, the intrinsic 2D light distribution of each star cluster is calculated by scaling the 2D PSF in the relevant photometric band to best match the observed 2D light distribution around the cluster. 

The PSFs of seeing limited observations are approximately Gaussian in shape and can be accurately characterised using their FWHM alone. However, the PSFs of diffraction limited observations have clear Airy ring patterns which cannot be accounted for with a pure Gaussian PSF model. Therefore, we use observed 2D light distributions of the tip-tilt star (in the NIRC2 filters) and isolated stars (in the HST images) as the PSF models for our photometry calculations. 

The PSF can vary significantly across the FOV when performing observations using adaptive optics. The turbulence sampled by the reference star is only an accurate representation of the turbulence along the line of sight to the target if the angular distance between the target and the reference star is less than the isokinetic angle $\theta_k$. This isokinetic angle is approximately 75 arcsec for the excellent conditions at Mauna Kea. Our tip-tilt star is only 17.6 arcseconds from the centre of MCG08, and therefore we would not expect to see any variation in PSF across our field. We do not observe any elongation of star clusters along the axis toward the tip-tilt star, and the radial emission profiles of the clusters do not appear to be dependent on the azimuthal orientation of the measurement. The absence of these common signatures confirms that the PSF variation over the FOV of our NIRC2 images is negligible and justifies our use of a single PSF.

\subsection{Cluster magnitudes}
\label{subsubsec:err}
We use the FASTPHOT procedure in IDL (an adapted version of the IRAF procedure DAOFIND) to extract the counts from each star cluster in each photometric band. We convert the counts to magnitudes using the published photometric zeropoints for the HST filters and the standard stars observed with NIRC2. The magnitude measurements for each star cluster are listed in Table \ref{Table:appendix} and the magnitude distributions are shown in Figure \ref{Fig:maghist}. The F814W and F435W magnitudes are primarily lower limits and therefore the intrinsic magnitude distributions are unknown.

The magnitude errors (calculated from variance in the sky level) are typically on the order of 0.1 mag (\mbox{$\sim$2\%}). However we must also consider errors introduced by our choice of PSF model. If the model PSF is sharper than the intrinsic PSF of the star clusters, then the calculated flux for each of the clusters will be lower than the intrinsic flux. We use the GETPSF procedure in IDL to calculate the approximate Gaussian FWHM of the PSF models and the star clusters. The model PSFs are sharper than the star cluster PSFs by up to a factor of 5. Smoothing the PSF models by a factor of 5 decreases the calculated Kp and J band magnitudes of the clusters by $\la \, 0.06$ and $\la \, 0.45$ respectively. This PSF characterisation error augments the negative error bars on the magnitudes but not the positive error bars, resulting in asymmetric errors.

\subsection{Flexible Stellar Population Synthesis models}
Stellar population synthesis (SPS) models bridge the gap between the observed spectral energy distributions (SEDs) of galaxies and their internal physical properties. The shape and normalization of the UV to IR spectra of star-forming galaxies are determined primarily by their star formation and chemical enrichment histories as well as the amount of dust attenuating the stellar light. Photometric and/or spectroscopic observations of galaxies over a wide wavelength range can therefore be used to constrain the properties of their stellar populations. 

We calculate model grids using the Flexible Stellar Population Synthesis (FSPS) models of \citet{Conroy10}. The Hertzsprung-Russell diagram is populated using spectra from the semi-empirical BaSeL3.1 stellar spectral library \citep{Lejeune97, Lejeune98, Westera02}. Isochrones are generated from the stellar evolution models of \citet{Marigo07} and \citet{Marigo08}, covering a range of initial masses (\mbox{0.15 $\textless \, M \, \textless$ 100 $\rm M_{\odot}$}), ages (\mbox{$10^{6.6} \textless \, t \, \textless 10^{10.2}$ yr}, \mbox{$\Delta$ (log t) = 0.05}) and metallicities (\mbox{$\rm 10^{-4} \, \textless \, Z \, \textless \, 0.030$}, \mbox{$\Delta$ Log(Z) = 0.1}). The integrated light of a simple stellar population (SSP) is calculated by summing stellar spectra along a single isochrone, weighted by stellar mass according to the chosen IMF and convolved with an appropriate SFH and dust attenuation prescription. The source-dust geometry configuration in MCG08 can be approximated as a point source attenuated by a foreground screen. The main source of opacity in young starburst galaxies is clumpy shells of dust embedded in HII regions. As the most massive stars evolve, stellar winds and supernova-driven outflows push gas and dust out of the HII regions \citep{Calzetti01}. 

We adopt a Salpeter IMF (consistent with previous works measuring the ages of young star clusters in merging systems, see e.g. \citealt{Pollack07, Wilson06, Whitmore02, Gilbert00}) and the \citet{Calzetti01} extinction law. (We note that adopting the Chabrier IMF does not alleviate the age-optical depth degeneracy preventing us from derving ages from the photometry and therefore does not change our results). We calculate models at solar metallicity (\mbox{Z = 0.0190}), with 120 optical depths ranging from \mbox{$\tau$ = 0 - 30} (\mbox{$\Delta \tau$ = 0.25}), and 188 ages spanning the full range covered by the stellar spectral libraries. FSPS produces UV-mm spectra (\mbox{91\AA \, $\leq \, \lambda \leq$ 1 cm}) as well as magnitudes in many common photometric filters (including F814W and F435W) for each of the 22560 models in the final grid. The NIRC2 filters are not included, so we use the magnitudes calculated for the UKIRT-WFCAM J band and TwoMass K band filters which have very similar transmission curves to the NIRC2 J and Kp band filters. 

\subsection{Constructing age probability distribution functions}
We construct an age probability distribution function (PDF) for each star cluster in our sample by assigning each model a weighting determined by the difference in the \mbox{J - Kp} colours of the cluster and the model. Models which are inconsistent with the measured \mbox{J - Kp} colours of the cluster (within the 1$\sigma$ error bars) or the limits on the \mbox{F814W - J} and \mbox{F435W - J} colours are given a weighting of zero. Models which are consistent with the observed photometry of the cluster are weighted under the assumption that the observed star cluster fluxes are drawn from a normal distribution (and therefore that the magnitudes are distributed log-normally). The PSF characterisation error (see Section \ref{subsec:psf_phot}) increases the probability density of fluxes above the mean relative to the standard log-normal distribution, which is accounted for by applying different standard deviations to the log-normal distribution on either side of the mean. Once constructed, the array of weights is normalized to a total sum of one and summed along the optical depth axis to create a one dimensional vector of weights as a function of star cluster age.

\begin{figure}
\centerline{\includegraphics[scale=0.5] {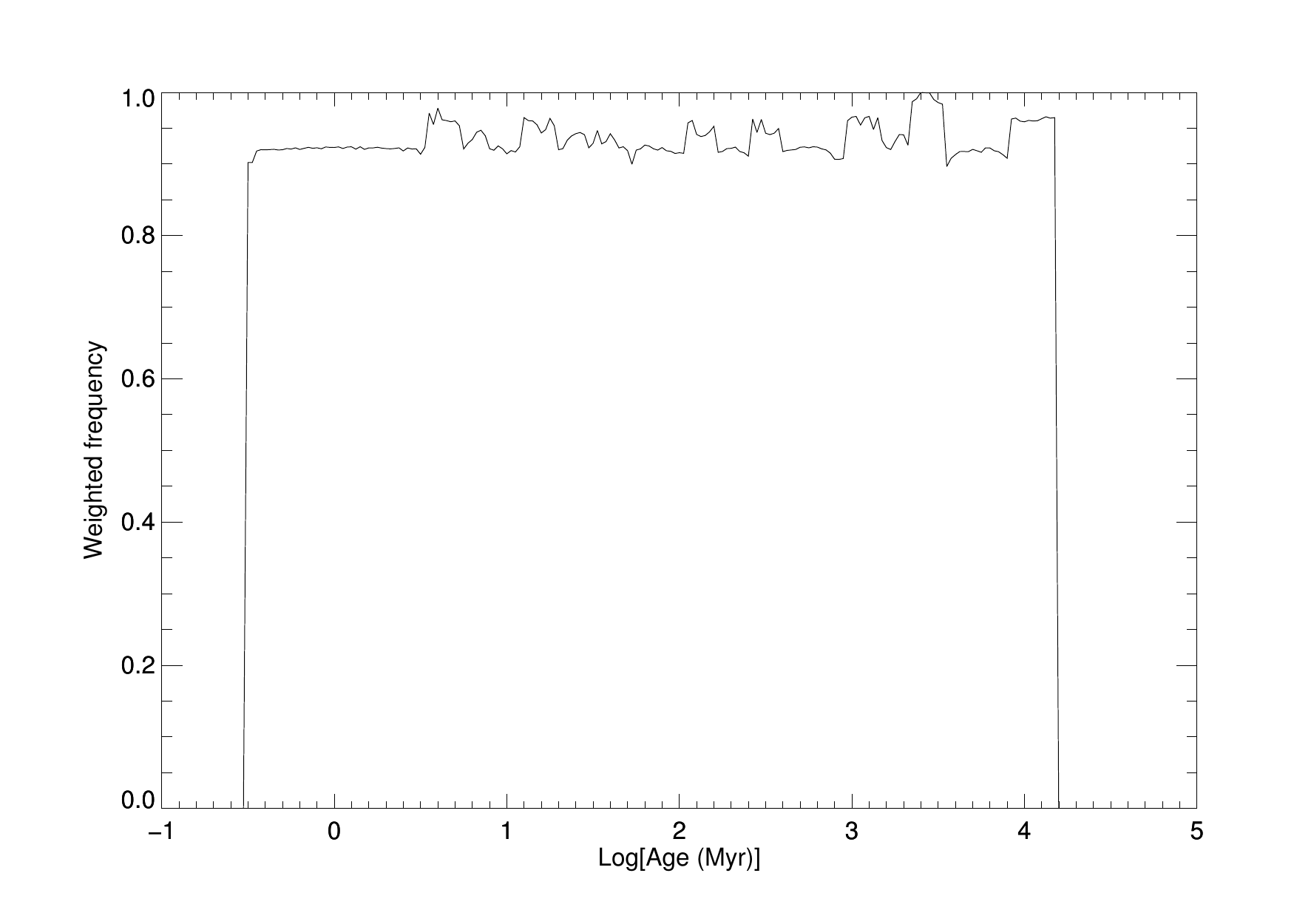}}
\caption{Example of an age probability distribution function constructed by selecting FSPS models consistent with the magnitude measurements and limits for a star cluster, and then weighting each of these models according to the difference between their J-Kp colour and the measured color of the cluster. The limits on the optical magnitudes are not sufficient to break the age-optical depth degeneracy.}
\label{Fig:age_PDF}
\end{figure}

Figure \ref{Fig:age_PDF} shows an example age PDF for one of the star clusters which is representative of the average sample properties. It is clear that our photometric information is not sufficient to constrain the ages of the star clusters in our sample. The F814W and F435W magnitude limits provide only limits on the optical depth, which are not strong enough to break the age - optical depth degeneracy. 

\begin{table*}
	\caption{Kp, J, F814W and F435W band apparent magnitudes measured from NIRC2 and HST imaging, Brackett decrements measured from the \hs\ OSIRIS data and implied optical depths for each of the 41 star clusters in our sample. Clusters which do not have listed Brackett decrements and optical depths are not covered by our OSIRIS observations.}	
	\begin{tabularx}{\textwidth}{RRRRRRR}
    \hline
    Identifier & \multicolumn{4}{c}{Apparent Magnitudes} & \Brg/\Brd\ & $\tau_{550}$ \\ \hline
   	~ & Kp & J & F814W & F435W & ~ & ~ \\ \hline 
1 & 19.8  & 20.9  & > 21.6 & > 25.2 & -  & - \\
2 & 17.9  & 20  & > 22.4 & > 26.9 & -  & - \\
3 & 19.5  & 21.1  & > 22.2 & > 26.2 & -  & - \\
4 & 17.7  & 19.9  & > 21.9 & > 27 & 2.45$\pm$     0.95  & 16.37$\pm$      32.89 \\
5 & 18.1  & 20.3  & > 22.1 & > 27.1 & -  & - \\
6 & 18.8  &  > 20.7  & > 22.8 & > 26.4 & -  & - \\
7 & 17.5  & 19.9  & > 22.8 & > 27.2 & 2.31$\pm$     0.53  & 14.33$\pm$      18.55 \\
8 & 18.3  & 20.6  & > 23 & > 26.7 & -  & - \\
9 & 17.6  & 19.8  & > 23 & > 27.6 & -  & - \\
10 & 17.7  &  > 19.8  & > 22.7 & > 27.3 & -  & - \\
11 & 16.8  & 19.8  & > 21.9 & > 26.6 & -  & - \\
12 & 17.2  & 19.6  & > 23.2 & > 27.6 & 1.91$\pm$     0.14  & 7.81$\pm$      4.88 \\
13 & 18.6  & 21  & > 23.4 & > 27.3 & -  & - \\
14 & 18.3  &  > 21.1  & > 23.2 & > 27.5 & -  & - \\
15 & 17.3  &  > 19.9  & > 22.3 & > 27.4 & 2.37$\pm$     0.16  & 15.25$\pm$      5.80 \\
16 & 18  &  > 21.1  & > 22 & > 27.7 & -  & - \\
17 & 16.9  & 19.5  & > 22.8 & > 27.3 & 1.62$\pm$     0.23  & 2.10$\pm$      8.10 \\
18 & 17.8  & 20.9  & > 22.3 & > 27.1 & -  & - \\
19 & 18.2  & 20.8  & > 23.5 & > 27.4 & -  & - \\
20 & 17.5  &  > 20.2  & > 22.1 & > 26.9 & 1.80$\pm$     0.13  & 5.72$\pm$      4.66 \\
21 & 18  &  > 20.6  & - & - & 2.07$\pm$     0.10  & 10.56$\pm$      3.57 \\
22 & 16.1  & 19  & > 23.1 & > 28.2 & 2.20$\pm$     0.20  & 12.73$\pm$      7.14 \\
23 & 16.8  &  > 19.6  & > 23.5 & > 28.4 & 2.12$\pm$     0.14  & 11.35$\pm$      5.07 \\
24 & 17.7  &  > 20.7  & > 21.2 & > 26.2 & 2.07$\pm$     0.10  & 10.56$\pm$      3.57 \\
25 & 16.8  & 19.6  & > 22.6 & > 27 & 1.96$\pm$    0.08  & 8.78$\pm$      2.76 \\
26 & 16  & 19.1  & > 22.1 & > 27 & 2.30$\pm$     0.11  & 14.18$\pm$      3.94 \\
27 & 16.3  & 19.4  & > 22.8 & > 27 & 2.60$\pm$    0.07  & 18.51$\pm$      2.69 \\
28 & 18.1  &  > 21.5  & > 23.3 & - & 2.77$\pm$     0.11  & 20.65$\pm$      3.81 \\
29 & 17.7  &  > 20.9  & > 22.1 & > 26.9 & 1.85$\pm$     0.13  & 6.76$\pm$      4.63 \\
30 & 17.7  & 20.6  & > 22.4 & > 27.1 & 1.61$\pm$     0.13  & 1.85$\pm$      4.79 \\
31 & 17.8  &  > 21.1  & > 22.2 & > 27.7 & 2.25$\pm$    0.08  & 13.54$\pm$      2.89 \\
32 & 18.4  & 21.1  & > 22.2 & > 26.9 & -  & - \\
33 & 17.7  &  > 20.7  & > 22.5 & > 27.1 & 1.90$\pm$     0.15  & 7.69$\pm$      5.26 \\
34 & 17.9  &  > 20.9  & > 22 & > 27 & 2.11$\pm$     0.13  & 11.24$\pm$      4.65 \\
35 & 17.6  & 20.3  & > 22.5 & > 26.8 & 2.15$\pm$     0.20  & 11.93$\pm$      7.20 \\
36 & 18  & 20.7  & > 22.3 & > 27.6 & -  & - \\
37 & 18.2  & 20.5  & > 22.3 & > 26.4 & -  & - \\
38 & 19  & 20.7  & > 21.9 & > 25.8 & -  & - \\
39 & 18.1  & 20.7  & > 21.7 & > 25.5 & -  & - \\
40 & 18.4  & 20.9  & > 22.6 & > 26.2 & -  & - \\
41 & 19.1  &  > 21.6  & > 22 & > 25.6 & -  & - \\ \hline
    \end{tabularx}
    \label{Table:appendix}
\end{table*}

\subsection{Constraints on the optical depth from the \Brg/\Brd\ ratio}
\label{sec:OSIRIS_tau}
We place further constraints on the optical depth along the line of sight to each star cluster using the \mbox{\Brg\ $\lambda$2.165 $\mu$m}/\mbox{\Brd\ $\lambda$1.944 $\mu$m} ratio (calculated from the OSIRIS \hs\ data). The unreddened intensity ratios between hydrogen recombination lines are determined by their transition probabilities which depend on the temperature and density of the line-emitting gas. The optical depth along the line of sight can be determined by comparing the measured and intrinsic \Brg/\Brd\ ratios. However the ratio of two infrared lines will only probe hot dust, providing a lower limit on the optical depth. 

We convert the measured \Brg/\Brd\ ratio in each spaxel to an optical depth value using the \citet{Fischera05} extinction curve with $R_V^A = 4.5$ (see \citealt{Vogt13} for a summary of the calculation method). The \citet{Fischera05} extinction curve is similar to the \citet{Calzetti01} empirical extinction curve for starburst galaxies ($R^A_V$ = 4.3), and assumes attenuation by a distant, turbulent, isothermal dust screen. We adopt an unreddened \Brg/\Brd\ ratio of 1.523, appropriate for Case B recombination in a nebula with an electron temperature of \mbox{10,000 - 20,000 K} and an electron density of \mbox{$\sim$100 $\rm cm^{-2}$} \citep{Osterbrock06}. The average optical depth along the line of sight to the nuclear region of MCG08 is \mbox{$\bar{\tau}_{550}$} = 12.1, corresponding to 13 magnitudes of extinction in V band. The optical depth values for each cluster are listed in Table \ref{Table:appendix}.

We construct new age PDFs, restricting the optical depth of the models to be above the lower boundary of the 1$\sigma$ confidence interval for each cluster. Unfortunately, the close wavelength proximity of the \Brg\ and \Brd\ lines introduces a large scaling factor in the conversion between the \Brg/\Brd\ ratio and optical depth, producing large uncertainties. Constraints from the \Brg/\Brd\ ratio ultimately do not improve on the photometric constraints explored in the previous section.

\end{document}